\documentclass[twocolumn,prd,amsmath,superscriptaddress]{revtex4}

\usepackage{graphicx}
\usepackage{amssymb}
\usepackage{dcolumn}
\usepackage{bm}
\usepackage{amsmath}

\newcommand{\Mpl}{M_\mathrm{Pl}}
\newcommand{\Veff}{V_\mathrm{eff}}

\newcommand{\bmat}{\beta_\mathrm{m}}
\newcommand{\bgam}{\beta_\gamma}
\newcommand{\rhom}{\rho_\mathrm{m}}
\newcommand{\rhog}{\rho_\gamma}
\newcommand{\meff}{m_\mathrm{eff}}
\newcommand{\Pgc}{{\mathcal P_\mathrm{pr}}}
\newcommand{\Pabs}{{\mathcal P_\mathrm{abs}}}
\newcommand{\tp}{\tau_\mathrm{pr}}

\newcommand{\fgam}{F_\gamma}
\newcommand{\faft}{F_\mathrm{aft}}
\newcommand{\ltot}{\ell_\mathrm{tot}}
\newcommand{\ltotone}{\ell_\mathrm{tot,1}}
\newcommand{\ltottwo}{\ell_\mathrm{tot,2}}
\newcommand{\Gdec}{\Gamma_{\mathrm{dec,}\gamma}}
\newcommand{\Gaft}{\Gamma_\mathrm{aft}}
\newcommand{\Agam}{{{\vec \Psi}_\gamma}}
\newcommand{\Aphi}{{{\Psi}_\phi}}

\newcommand{\dB}{{\vec {\delta B}}}
\newcommand{\dphi}{{\delta \phi}}

\newcommand{\xhat}{{\hat{x}}}
\newcommand{\yhat}{{\hat{y}}}
\newcommand{\zhat}{{\hat{z}}}
\newcommand{\khat}{{\hat{k}}}
\newcommand{\Pcg}{{\mathcal P}_{\phi \leftrightarrow \gamma}}
\newcommand{\fref}{f_\mathrm{ref}}
\newcommand{\fabs}{f_\mathrm{abs}}
\newcommand{\Aref}{A_\mathrm{ref}}
\newcommand{\xiref}{\xi_\mathrm{ref}}
\newcommand{\xiprop}{\xi_\mathrm{M}}
\newcommand{\eL}{\xi_\mathrm{L}}
\newcommand{\eR}{\xi_\mathrm{R}}
\newcommand{\sL}{s_\mathrm{L}}
\newcommand{\sM}{s_\mathrm{M}}
\newcommand{\sR}{s_\mathrm{R}}
\newcommand{\nL}{{n_\mathrm{L}}}
\newcommand{\nR}{{n_\mathrm{R}}}
\newcommand{\fgeom}{f_\mathrm{geom}}
\newcommand{\sfrag}{\sigma_\mathrm{frag}}
\newcommand{\Gfrag}{\Gamma_\mathrm{frag}}
\newcommand{\Gfm}{\Gamma_\mathrm{frag}^\mathrm{(max)}}
\newcommand{\Mlam}{M_\Lambda}
\newcommand{\phimin}{\phi_\mathrm{min}}

\begin{document} 

\title{Constraining chameleon field theories using the GammeV afterglow experiments}

\author{A. Upadhye}
\affiliation{Kavli Institute for Cosmological Physics, Enrico Fermi Institute, University of Chicago, Chicago, IL 60637}

\author{J.~H. Steffen}
\affiliation{Fermi National Accelerator Laboratory, PO Box 500, Batavia, IL 60510}

\author{A. Weltman}
\affiliation{Department of Applied Mathematics and Theoretical Physics, Cambridge CB2 0WA, United Kingdom}
\affiliation{Cosmology and Gravity Group, University of Cape Town, Rondebosch, Private Bag, 7700 South Africa}

\date{\today}

\begin{abstract}
The GammeV experiment has constrained the couplings of chameleon scalar fields to matter and photons.  Here we present a detailed calculation of the chameleon afterglow rate underlying these constraints.  The dependence of GammeV constraints on various assumptions in the calculation is studied.  We discuss GammeV--CHASE, a second-generation GammeV experiment, which will improve upon GammeV in several major ways.  Using our calculation of the chameleon afterglow rate, we forecast model-independent constraints achievable by GammeV--CHASE.  We then apply these constraints to a variety of chameleon models, including quartic chameleons and chameleon dark energy models.  The new experiment will be able to probe a large region of parameter space that is beyond the reach of current tests, such as fifth force searches, constraints on the dimming of distant astrophysical objects, and bounds on the variation of the fine structure constant.
\end{abstract}

\maketitle


\section{Introduction} \label{sec:intro}

Increasingly strong evidence has emerged over the past decade that the expansion of  the universe is accelerating, a phenomenon which can be explained by a scalar field ``dark energy'' with negative pressure~\cite{Komatsu_etal_2009,Dunkley_etal_2009,Kowalski_etal_2008,Percival_etal_2009,Ratra_Peebles_1988,Peebles_Ratra_1988,Caldwell_Dave_Steinhardt_1998}.  Couplings between such a scalar and Standard Model particles could lead to fifth forces observable at laboratory or solar system scales, that must be hidden in order for the dark energy to satisfy constraints on the non-observation of such forces.  There are three known ways to hide dark energy-mediated fifth forces: weak couplings between dark energy and matter~\cite{Wilczek_1982,Frieman_Hill_Stebbins_Waga_1995,Kim_Nilles_2009}; effectively weak couplings locally, as in the Dvali-Gabadadze-Porrati (DGP) brane world model and its generalizations~\cite{Dvali_Gabadadze_Porrati_2000,deRham_etal_2008,deRham_Hofmann_Khoury_Tolley_2008}; and an effectively large mass locally, as in chameleon theories~\cite{Khoury_Weltman_2004a,Khoury_Weltman_2004b,Brax_etal_2004}.  

The chameleon mechanism makes the effective mass of a scalar field grow substantially as the density of the surrounding matter is increased.  Chameleon theories are a particularly well-studied class of dark energy models~\cite{Carlson_Garretson_1994,Burrage_Davis_Shaw_2009,Brax_etal_2009,Brax_etal_2009b,Brax_etal_2007,Brax_etal_2007b,Chou_etal_2009,Ahlers_etal_2008,Gies_Mota_Shaw_2008,Mirizzi_Redondo_Sigl_2009}.  Since fifth forces at least as strong as gravity are already ruled out from solar system scales to submillimeter scales, the chameleon mechanism must raise the effective field mass above about $(0.1\textrm{ mm})^{-1} \sim 10^{-3}$~eV locally.   In addition to chameleon dark energies, the ``scalaron'' field, resulting from the transformation to Einstein frame of an $f(R)$ gravity theory, is able to satisfy local and stellar tests of gravity only through the chameleon mechanism~\cite{Hu_Sawicki_2007,Babichev_Langlois_2009,Upadhye_Hu_2009,Babichev_Langlois_2009b}.  

It was shown in \cite{Mota_Shaw_2007,Mota_Shaw_2006} that chameleons which couple strongly to matter have correspondingly large chameleon effects, allowing them to evade laboratory fifth force constraints such as \cite{Adelberger2007} on matter couplings $\bmat \sim 1$.  If these chameleons also couple strongly to photons, then they are ideally suited to probes of the afterglow phenomenon, the first of which was the GammeV experiment at Fermilab~\cite{Chou_etal_2009}.  Consider a closed, evacuated cylindrical chamber, with glass windows at the ends and a magnetic field in the interior.  Photons streamed through the windows will occasionally oscillate into chameleon particles in the background magnetic field.  If the mass of one of these chameleons in the walls of the chamber is greater than its total energy inside the chamber, then it will reflect from the wall; such a chameleon will be trapped in the chamber.  After the photon source has been turned off, any remaining chameleons will oscillate back into photons in the magnetic field, producing an observable ``afterglow'' of photons.

 Here, we study the oscillation between photons and chameleon particles in the cylindrical vacuum chambers used in afterglow experiments such as GammeV.  We compute the afterglow and decay rates per chameleon particle as a function of the dimensionless chameleon-photon coupling constant $\bgam$.  Our calculation allows for absorptivity in the chamber walls, chameleon-photon phase shifts due to a nonzero chameleon mass $\meff(\textrm{chamber})$, and additional phase shifts $\xiref$ due to reflections from chamber walls, as found in \cite{Brax_etal_2007}.  For a particle in a superposition of chameleon and photon states, the glass windows at the ends of the chamber act as quantum measurement devices; photons are transmitted, while chameleons are reflected.  The chameleon decay and afterglow rates can be computed by keeping track of the decline in chameleon amplitude between successive quantum measurements by the windows.

Armed with these rates, we proceed to compute the expected afterglow flux as a function of time in GammeV, as well as the resulting constraints on the chameleon-photon coupling as a function of the chameleon mass, a result that was discussed in ref.~\cite{Chou_etal_2009}.  We then forecast the constraints that will be attained by an upcoming experiment, GammeV--CHASE, expected to take data in the winter of 2009-2010.  GammeV--CHASE was designed to improve in several ways upon the first GammeV chameleon experiment, hereafter referred to as GammeV.  We show that GammeV--CHASE will improve constraints on large photon couplings $\bgam$ by several orders of magnitude, bridging the gap between GammeV constraints and the collider constraints of ref.~\cite{Brax_etal_2009}.  Furthermore, it will reach chameleon masses several times greater than those excluded by GammeV, allowing us to probe masses at the dark energy scale $\rho_\mathrm{de}^{1/4} = 2.4\times 10^{-3}$~eV.  

Finally, we use our afterglow computation to forecast GammeV--CHASE constraints on specific chameleon models.  We begin with power law potentials, devoting most of our attention to $\phi^4$ potentials, which are well-understood and ubiquitous in particle physics.  GammeV--CHASE constraints on $\phi^4$ chameleons will complement the fifth force experiments of E\"ot-Wash \cite{Adelberger2007} by probing strong matter couplings.  Next, we study two different models of chameleon dark energy, with inverse power law potentials and exponential potentials.  GammeV--CHASE will probe large areas of the $\bmat$,~$\bgam$ parameter space of each of these dark energy models. These parameter regions are largely unexplored by the current data, including constraints on the dark energy equation of state, variations in the electromagnetic fine structure constant, and the dimming of distant objects through photon-scalar oscillation in the Galactic magnetic field.  Thus, the GammeV--CHASE laboratory search for afterglow will complement current cosmological probes of dark energy, vastly extending our constraints on dark energy couplings to matter and photons.

The paper is organized as follows.  After a brief discussion of chameleon physics in Sec.~\ref{sec:chameleons}, we study chameleon-photon conversion in afterglow experiments in Sec.~\ref{sec:afterglow}, and compute the decay and afterglow rates for GammeV and GammeV--CHASE in Sec.~\ref{sec:gammev}.  We constrain chameleons using GammeV and forecast constraints for GammeV--CHASE in 
Sec.~\ref{sec:constraints}, before concluding in Sec.~\ref{sec:conclusion}.


\section{Chameleon field theories} 
\label{sec:chameleons}

\subsection{Equations of motion}

We consider chameleons with action
\begin{eqnarray}
S 
&=& 
\int d^4x \sqrt{-g}{\Big (}
\frac{1}{2}\Mpl^2R
-
\frac{1}{2}\partial_\mu\phi\partial^\mu\phi 
- V(\phi) \nonumber\\
&&- \frac{1}{4}e^{\bgam \phi/\Mpl}F^{\mu\nu}F_{\mu\nu}
+ {\mathcal L}_\mathrm{m}(e^{2\bmat\phi/\Mpl}g_{\mu\nu},\psi^i_\mathrm{m}) {\Big )}\quad
\label{e:action}
\end{eqnarray}
where $\bgam$ and $\bmat$ are, respectively, the dimensionless chameleon couplings to photons and matter.  We note that elsewhere in the literature, these couplings are expessed in dimensional form as $g_\gamma \equiv \bgam/\Mpl$ and $g_\mathrm{m} \equiv \bmat/\Mpl$; for a typical value $\bgam = 10^{12}$ probed by GammeV~\cite{Chou_etal_2009}, we have $g_\gamma = 4.1\times 10^{-7}\textrm{ GeV}^{-1}$.  ${\mathcal L}_\mathrm{m}$ is the matter Lagrangian; the matter coupling $\bmat$ is assumed to be universal to all species of matter.  Varying with respect to $\phi$ and the electromagnetic field, we find the equations of motion in the presence of a constant matter density $\rhom$,
\begin{eqnarray}
\Box \phi &=& -\frac{\partial \Veff}{\partial \phi}\\
\Veff(\vec x,\phi) &=& V(\phi) 
+ e^\frac{\bmat \phi}{\Mpl}\rhom 
+ \frac{1}{4}e^{\frac{\bgam \phi}{\Mpl}}F^{\mu\nu}F_{\mu\nu}  \\
\partial_\mu (e^\frac{\bgam \phi}{\Mpl}F^{\mu\nu}) &=& 0
\end{eqnarray}
with the other two of Maxwell's equations unchanged.  That is, the field couples to the matter density $\rhom$ and the electromagnetic field Lagrangian density $\rhog = (F^{\mu\nu}F_{\mu\nu})/4 = (|\vec B|^2 - |\vec E|^2)/2$.  Perturbing about a constant background scalar $\phi$ and magnetic field $\vec B$, we see that oscillations from chameleon to photon are described by
\begin{equation}
\Box \, \dB 
+ \frac{\bgam}{\Mpl} {\vec \nabla} \times (\vec B \times \vec \nabla) \dphi
= 0
\label{e:phigam}
\end{equation}
to first order in the perturbations~\cite{Raffelt_Stodolsky_1988}.

\subsection{Chameleon effects} 

The field value $\phimin$ at the minimum of the effective potential, and, hence, the effective mass $\meff^2 = V_{\mathrm{eff},\phi\phi}(\phimin)$, can vary with the background matter density and electromagnetic fields.  If $\meff$ increases significantly with $\rhom$ or $\rhog$, then the field is called a chameleon field.  By acquiring a large mass at laboratory and solar system densities, the chameleon can ``hide'' from fifth force constraints on ordinary matter-coupled scalars.  As a simple example, consider a power law chameleon, $V(\phi) = g \phi^{\mathcal N}$, in a background with negligible $\vec E$ and $\vec B$.  Assuming that $\bmat \phimin/\Mpl \ll 1$, which is required by current constraints, the mass scales as $\meff \propto \rhom^\eta$ with $\eta = ({\mathcal N}-2)/(2{\mathcal N}-2)$.  If ${\mathcal N}=2$, that is, the potential is just a mass term, then $\eta=0$; $\meff$ does not scale with density, and the field is not a chameleon.  For ${\mathcal N}=4$, on the other hand, $\eta = 1/3$, and references \cite{Mota_Shaw_2007,Mota_Shaw_2006} show that, for sufficiently large $\bmat$, chameleon effects allow this theory to evade fifth force constraints from torsion pendulum experiments such as E\"ot-Wash~\cite{Adelberger2007}.  Chameleon fields strongly coupled to matter acquire large masses inside the layer of foil used to keep the source and test masses electrostatically isolated.  If the chameleon Compton wavelength $\meff^{-1}$ is much less than the thickness of the foil, then chameleon effects in the foil will screen the fifth force of the source mass on the test mass, rendering the chameleon field undetectable.  


\section{Chameleon-photon conversion in afterglow experiments}
{\label{sec:afterglow}}

\subsection{Photon production}

Let the chameleon and photon amplitudes be $\Aphi = \dphi$ and $\Agam = \dB/k$, respectively.  We consider the oscillation into photons of chameleons that are already trapped inside the GammeV chamber for the case of a small mixing angle.  Thus we assume that $|\Agam| = 0$ initially, and that $|\Agam| \ll |\Aphi| \approx 1$ throughout the calculation.  In the presence of a constant magnetic field in the $\xhat$ direction, ${\vec B} = B\xhat$, (\ref{e:phigam}) becomes
\begin{equation}
\left(-\frac{\partial^2}{\partial t^2} - {\vec k}^2\right)\Agam
=
\frac{k \bgam B}{\Mpl} \khat \times (\xhat \times \khat) \Aphi.
\label{e:wave}
\end{equation}
In the relativistic limit, $(-\partial^2/\partial t^2 - |\vec k|^2) = (i \partial/\partial t - k)(i\partial /\partial t + k) \approx 2k (i\partial/\partial t - k)$, this has the solution
\begin{eqnarray}
\Agam(t) 
&=&
-ie^{-ikt} \frac{\bgam B}{2\Mpl} 
\int_0^t \vec{a}(\hat k) \exp\left(-\frac{i \meff^2 t'}{2k}\right) dt'
\label{e:psig_int}\\
&=&
-i 
e^{-ikt - \frac{i \meff^2 t}{4 k}}
\frac{2 k \bgam B}{\meff^2 \Mpl}
\sin\left(\frac{\meff^2 t}{4k}\right)
{\vec a}(\hat k),
\label{e:psig_segment}
\end{eqnarray}
where $\meff$ is the effective mass of the chameleon, ${\vec k} = k\khat = k(\sin(\theta)\cos(\varphi)\xhat + \sin(\theta)\sin(\varphi)\yhat + \cos(\theta)\zhat)$ is the particle momentum, and ${\vec a}(\khat) \equiv \khat \times (\xhat \times \khat) = (1-\sin^2(\theta)\cos^2(\varphi))\xhat - \sin^2(\theta)\sin(\varphi)\cos(\varphi)\yhat - \sin(\theta)\cos(\theta)\cos(\varphi)\zhat$.   Using $|\vec a|^2 = \sin^2(\varphi) + \cos^2(\theta)\cos^2(\varphi)$, and setting $\theta=0$, we recover the familiar formula for the oscillation probability for a path perpendicular to $\vec B$,
\begin{equation}
\Pcg = |\Agam|^2 = C^2 \sin^2\left(\frac{\meff^2 t}{4k}\right),
\label{e:Pcg}
\end{equation}
where we have defined $C = 2 k \bgam B / (\Mpl \meff^2)$.

Next, assume that the chameleon particle is incident upon a wall, with normal vector $\hat n$.  We consider chameleon models for which the effective mass inside the wall is much greater than the chameleon energy $\omega = \sqrt{k^2 + \meff^2}$ inside the chamber.  Thus the chameleon cannot penetrate the wall, and tunneling is negligible; the particle must bounce. The bounce has three effects:
\begin{enumerate}
\item the direction changes, $\hat k \rightarrow \hat k - 2(\hat k \cdot \hat n)\hat n$, with a corresponding change in $\vec a$;
\item the photon component has a probability $\fabs = 1 - \fref$ of being absorbed in the walls;
\item the photon is phase shifted by an angle $\xiref$ relative to the chameleon \cite{Brax_etal_2007}.
\end{enumerate}
These last two imply that $\Agam \rightarrow \Aref \Agam$ due to a bounce, where $\Aref \equiv \fref^{1/2} e^{i\xiref}$.

\subsection{Afterglow experiments}

A simple afterglow experiment will trap chameleon particles in a cylindrical chamber, of radius $R$, with glass windows at the entrance and exit.  A  magnetic field region of length $L$ inside the cylinder will be offset from the entrance by a length $\ell_1$, and from the exit by a length $\ell_2$, for a total chamber length of $\ell_\mathrm{tot} = \ell_1 + \ell_2 + L$.  Photons stream into the chamber through the entrance window, and out through the exit, oscillating into chameleon particles with a probability given by (\ref{e:Pcg}).  For a particle in a superposition of chameleon and photon states, a glass window serves as a quantum measurement device; a photon will pass through, while a chameleon will bounce.  Thus we can assume that any particle bouncing from either window is in a pure chameleon state.  After the photon source is turned off, a photomultiplier tube (PMT) in a dark box outside the exit window will begin to look for afterglow photons produced by chameleons trapped in the chamber.  

Let the origin of the coordinate system be at the center of the entrance window, with the cylinder extending in the positive $z$ direction, and the magnetic field in the $x$ direction.  Assume that the particle begins at the origin; we will test the effects of this assumption later.  Let $\theta$ and $\varphi$ be defined in the usual way; $\theta$ is the angle between $\hat k$ and the $z$ axis, and $\varphi$ is the angle between $\hat k - (\hat k \cdot \hat z)\hat z$ and the $x$ axis, where $\hat k$ is the initial direction of the particle.  For small $\theta$, the particle will not bounce inside the field region, and (\ref{e:Pcg}) suffices to compute the probability of photon production.

\begin{figure}[tb]
\begin{center}
\includegraphics[width=3.3in]{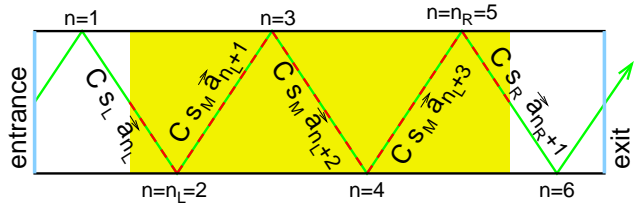}
\caption{Particle in an afterglow experiment.  The magnetic field region is shaded.  Inside this region, where chameleon-photon oscillation takes place, the particle trajectory is drawn as a dashed path.  Next to each dashed segment is shown the contribution of that segment to the total photon amplitude for the case where $\fref=1$, $\xiref=0$, and $\meff \ll 4\pi k / L$. \label{f:Psi_gamma}}
\end{center}
\end{figure}

At larger $\theta$, the particle will bounce inside the magnetic field region.  This case is sufficiently complicated that we begin with a simple example in which the chamber walls are perfectly reflective, the phase shift associated with wall reflection is zero, and the chameleon mass is small enough that phase differences between the chameleon and photon as they propagate through the chamber may be neglected.  Figure~\ref{f:Psi_gamma} shows a sample particle trajectory.  The particle begins in a pure chameleon state at the entrance window, shown at the left end of the chamber, and heads to the right, towards the magnetic field.    Let $z_1, z_2, \ldots, z_N$ be the sequence of $z$ values at which the particle bounces from a wall.  For a particle beginning at the origin, $z_n = (n-1/2)\Delta z$, with $\Delta z = 2R \cot(\theta)$.  Let the leftmost and rightmost bounces inside the magnetic field be $\nL$ and $\nR$, respectively; the total number of bounces in the B field region is then $n_B = \nR - \nL + 1$.

The sample trajectory shown in Fig.~\ref{f:Psi_gamma} has six bounces, four of which are in the B field region, and is made up of seven segments, five of which have some overlap with the B field region.  In the figure, a dashed line denotes that portion of a trajectory which is in the B field region, and, hence, contributes to the photon amplitude.  The contribution of each segment is given by (\ref{e:psig_segment}).  In our approximation that $\meff$ is small, the chameleon and photon are in phase; $\Aphi,\: \Agam \propto \exp({-ikt})$.  Thus we can neglect the phase factor in front of $\Agam$, and write the contribution of the $n$th segment as $C \sin(\meff^2 t/(4k)) {\vec a}_n$, where $t$ is the time spent inside the B field region, $\vec a_n \equiv \vec a(\hat k_n)$, and $\hat k_n$ is the direction of the particle before the bounce at $z_n$.  Furthermore, $C \sin(\meff^2 t/(4k)) \approx \bgam B t / (2\Mpl)$ at low $\meff$, so the contribution of each segment is proportional to its length inside the magnetic field region.

The leftmost segment in the B field region, the segment between bounces $\nL-1$ and $\nL$ (in Fig.~\ref{f:Psi_gamma}, bounces $1$ and $2$) is only partially inside the magnetic field, so $t_\mathrm{L} = (z_\nL - \ell_1) \sec(\theta)$ for that segment.  Similarly, the rightmost segment in the magnetic field, between bounces $\nR$ and $\nR+1$ (here, $5$ and $6$), has $t_\mathrm{R} = (\ell_1 + L - z_\nR) \sec(\theta)$.  Meanwhile, all of the segments in the middle, $\nL < n \leq \nR$, spend an equal amount of time in the magnetic field, $t_\mathrm{M} = \Delta z \sec(\theta)$.  Defining 
\begin{eqnarray}
\sL
&=&
\sin\left(\frac{\meff^2(z_\nL-\ell_1)}{4k\cos(\theta)}\right)
\\
\sM
&=&
\sin\left(\frac{\meff^2\Delta z}{4k\cos(\theta)}\right)
= \sin\left(\frac{\meff^2 R}{2k\sin(\theta)}\right)
\\
\sR
&=&
\sin\left(\frac{\meff^2 (\ell_1 + L -z_\nR)}{4k\cos(\theta)}\right)
\end{eqnarray}
we see that each of the the leftmost, middle, and rightmost segments contributes $C \sL {\vec a}_\nL$, $C \sM {\vec a}_n$, and $C \sR {\vec a}_{\nR+1}$, respectively, where $\nL < n \leq \nR$.  The photon amplitude at the exit $z = \ell_B = \ell_1 + L$ of the B field region is simply the sum of these contributions, $\Agam(\ell_B) = C\sL {\vec a}_\nL + \sum_n C \sM {\vec a}_n + C \sR {\vec a}_{\nR+1}$, up to a complex phase factor which disappears when $\Agam$ is squared.

At this point, our choice of initial conditions simplifies the problem.  Recall that we have assumed the initial particle position to be the center of the entrance window.  The trajectory of such a particle will remain in the plane spanned by the $z$ axis and the initial direction $\hat k$, even after the particle bounces from the chamber walls.  This is because the particle momentum has only a radial component $\propto x \xhat + y \yhat$ and a component proportional to $\zhat$.  Each bounce reverses the sign of the radial component while leaving the $\zhat$ component unchanged.  Furthermore, each bounce reverses the sign of the $\zhat$ component of $\vec a$ while leaving the $\xhat$ and $\yhat$ components unchanged; for all odd $n$, $\vec a_n = \vec a_1$, and for all even $n$, $\vec a_n = \vec a_2$, with $\vec a_2 = \vec a_1 - 2(\vec a_1 \cdot \zhat)\zhat$.  The summation over segments inside the B field is considerably simplified, $\sum_n C \sM \vec a_n = C \sM \sum_n (a_{1,x}\xhat + a_{1,y}\yhat + (-1)^n a_{1,z}\zhat)$.

Next, we consider two complications to the problem: the possibility of photon absorption in the chamber walls, and the phase shift between chameleons and photons due to a bounce from the walls.  Even a polished metal surface will not be perfectly reflective; the fraction of incident photons reflected back into the chamber will be around $\fref \sim 0.9$.  Furthermore, since chameleons and photons bounce at slightly different distances from the wall, the bounce can introduce a nonzero phase shift $\xiref$ \cite{Brax_etal_2007}.  We model absorption in the walls by multiplying the photon amplitude by $\Aref = \fref^{1/2}e^{i\xiref}$ at each bounce.  Thus the photon amplitude just before the $\nL$th bounce is $\Agam(z_\nL^-) = C \sL \vec a_\nL$, and the amplitude just after that bounce is $\Agam(z_\nL^+) = \Aref C \sL \vec a_\nL$.  Before and after the next bounce, we have $\Agam(z_{\nL+1}^-) = \Aref C \sL \vec a_\nL + C \sM \vec a_{\nL+1}$ and $\Agam(z_{\nL+1}^+)= \Aref^2 C \sL \vec a_\nL + \Aref C \sM \vec a_{\nL+1}$, respectively, where we have assumed that $\nL+1 \leq \nR$.  Summing over all of the bounces, we find the photon amplitude at the exit of the B region, 
\begin{eqnarray}
\Agam^\mathrm{(low\textrm{ }mass)}(\ell_B) 
&=& 
\Aref^{n_B} C \sL \vec a_\nL 
+ 
C \sM \sum_n \Aref^{\nR+1-n} \vec a_n 
\nonumber\\
&&
+ 
C \sR a_{\nR+1},
 \end{eqnarray}
where $n_B = \nR - \nL + 1$ is the total number of bounces in the B region.

The final effect which we have neglected thus far is the phase shift associated with the chameleon-photon mass difference.  Eq.~(\ref{e:psig_segment}) implies that contributions to $\Agam$ from different segments of the path, which occur at different times, will have relative phases.  As before, the leftmost and rightmost segments in the B region contribute different amounts to the phase shift, since they are only partially inside the B region, while the middle segments all contribute the same phase shift.  Defining
\begin{eqnarray}
\xiprop 
&=& 
\frac{\meff^2\Delta z}{2k\cos(\theta)}
= \frac{\meff^2 R}{k\sin(\theta)} 
\\
A
&=&
\Aref e^{i\xiprop}
\\
\eL
&=&
\frac{\meff^2}{4k\cos(\theta)}
\left(\left(\nL-\frac{3}{2}\right)\Delta z -\ell_1\right)
\\
\eR
&=&
\frac{\meff^2}{4k\cos(\theta)}
\left(\left(\nR+\frac{1}{2}\right)\Delta z - \ell_B\right).
\end{eqnarray}
we have our final expression for the photon amplitude at the exit of the B region,
\begin{eqnarray}
\Agam(\ell_B)
&=&
-i C \exp\left(-\frac{ik\ell_B}{\cos(\theta)} - i \nR \xiprop\right)
\Bigg[
A^{n_B}e^{i\eL}\sL\vec a_\nL
\nonumber\\
&&
+ \sum_{j=1}^{n_B-1} A^{n_B-j} \sM \vec a_{\nL+j}
+ e^{i\eR} \sR \vec a_{\nR+1}
\Bigg], 
\label{e:psi_Bexit}
\end{eqnarray}
where we have simplified our notation by combining $\Aref$ and $\exp(i\xiprop)$ into $A$.

As above, our choice of initial conditions allows us to compute explicitly the sum in (\ref{e:psi_Bexit}),
\begin{eqnarray}
\sum_{j=1}^{n_B-1} A^{n_B-j} \vec a_{\nL+j}
&=&
(a_{1,x}\xhat + a_{1,y}\yhat)\frac{A-A^{n_B}}{1-A}
\nonumber\\
&\quad&
+ a_{1,z}\zhat (-1)^\nR \frac{(-A) - (-A)^{n_B}}{1+A}.\qquad
\label{e:Psi_sum}
\end{eqnarray}

After the particle leaves the magnetic field region, the photon and chameleon states decouple, and no further oscillation occurs.  Thus, the probablity that the particle will be a photon, when a quantum measurement is made by the exit window, is $|\Agam(\ell_B)|^2 \fref^{N-\nR}$, where $N-\nR$ is the number of times that the particle bounces after leaving the field region.

Since we also want to compute the decay rate of the chameleon, we must keep track of the photons lost to absorption in the chamber walls, inside the magnetic field region.  At each bounce, $\Agam \rightarrow \Aref \Agam$, the absorption probability is incremented by $(1-\fref)|\Agam|^2 = \fabs|\Agam|^2$; since the tunneling rate is negligible, the sum of the reflection and absorption probabilities must be equal.  We find the total absorption probability $\Pabs$ by summing $\fabs |\Agam|^2$ over all of the bounces in the field region.
\begin{eqnarray}
\Pabs
&=&
\fabs C^2 \Bigg[
\sL^2 \sum_{j=0}^{n_B-1} \fref^j |\vec a_\nL|^2
\nonumber\\
&&
+
\sM^2 \sum_{j=1}^{n_B-1} 
\left|\sum_{\ell=1}^j A^{j-\ell}\vec a_{\nL+\ell}\right|^2
\nonumber\\
&&
+ \sL\sM \sum_{j=1}^{n_B-1} \sum_{\ell=1}^j
\vec a_\nL \cdot \vec a_{\nL+\ell}
\nonumber\\
&&\qquad \times
\left(e^{i\eL}A^j (A^*)^{j-\ell} + e^{-i\eL}(A^*)^j A^{j-\ell}\right)
\Bigg]\qquad
\label{e:Pabs}
\end{eqnarray}
Once again, all of the summations can be carried out explicitly for particles originating at the center of the entrance window:
\begin{eqnarray}
\sum_{j=0}^{n_B-1} \fref^j |\vec a_\nL|^2 
=
|\vec a_1|^2 \frac{1-\fref^{n_B}}{1-\fref}&&
\qquad\qquad\qquad\qquad\quad\,\,\,
\label{e:Pabs_sum1}
\end{eqnarray}
\begin{eqnarray}
&&
\sum_{j=1}^{n_B-1} 
\left|\sum_{\ell=1}^j A^{j-\ell}\vec a_{\nL+\ell}\right|^2 
\nonumber\\
&&\qquad =
\frac{a_{1,x}^2+a_{1,y}^2}{(1-A)(1-A^*)}
\Bigg[n_B-1 + \frac{\fref-\fref^{n_B}}{1-\fref}
\nonumber\\
&&\qquad\qquad
- 
\left(\frac{A-A^{n_B}}{1-A} + \frac{A^*-(A^*)^{n_B}}{1-A^*}\right)
\Bigg]
\nonumber\\
&&\qquad\quad
+ \frac{a_{1,z}^2}{(1+A)(1+A^*)} 
\Bigg[n_B-1 + \frac{\fref-\fref^{n_B}}{1-\fref}
\nonumber\\
&&\qquad\qquad
- 
\left(\frac{-A-(-A)^{n_B}}{1+A} + \frac{-A^*-(-A^*)^{n_B}}{1+A^*}\right)
\Bigg]\qquad
\label{e:Pabs_sum2}
\end{eqnarray}
\begin{eqnarray}
&&
\sum_{j=1}^{n_B-1} \sum_{\ell=1}^j
\vec a_\nL \cdot \vec a_{\nL+\ell}
\nonumber\\
&&\qquad =
\frac{a_{1,x}^2 - a_{1,y}^2}{1-A}
\left(\frac{A^*-(A^*)^{n_B}}{1-A^*} - \frac{\fref-\fref^{n_B}}{1-\fref}\right)
e^{-i\eL}
\nonumber\\
&&\quad\qquad 
+
\frac{a_{1,z}^2}{1+A}
\left(\frac{-A^*-(-A^*)^{n_B}}{1+A^*} -\frac{\fref-\fref^{n_B}}{1-\fref}\right)
e^{-i\eL}
\nonumber\\
&&\quad\qquad
+
\textrm{ complex conjugate}.
\label{e:Pabs_sum3}
\end{eqnarray}

A chameleon particle with initial direction specified by $\theta$ and $\varphi$  traverses the chamber in a time $\ltot \sec(\theta)$.  It has a probability $\Pabs + |\Agam(\ell_B)|^2$ of producing a photon, and a probability $|\Agam(\ell_B)|^2 \fref^{N-\nR}$ of producing a photon that escapes through the exit window.  Thus the contribution of this $\theta$ and $\varphi$ to the decay rate is the photon production probability per unit time, $(\Pabs + |\Agam(\ell_B)|^2)/(\ltot \sec(\theta))$, and similarly for the afterglow rate.  We find the total decay and afterglow rates per particle by integrating over angles.
\begin{eqnarray}
\Gdec
&=&
\frac{1}{2\pi}
\int_0^{\pi/2} \sin(\theta)d\theta
\int_0^{2\pi} d\varphi
\nonumber\\
&&\qquad
\left(\Pabs(\theta,\varphi) 
+ \left|\Agam(\ell_B,\theta,\varphi)\right|^2\right)
\frac{\cos(\theta)}{\ltot}\qquad
\label{e:Gdec}
\\
\Gaft
&=&
\frac{1}{4\pi}
\int_0^{\pi/2} \sin(\theta)d\theta
\int_0^{2\pi} d\varphi
\nonumber\\
&&\qquad
\fref^{N(\theta)-\nR(\theta)}
\left|\Agam(\ell_B,\theta,\varphi) \right|^2 
\frac{\cos(\theta)}{\ltot}\qquad
\label{e:Gaft}
\end{eqnarray}
The extra factor of two in the decay rate accounts for chameleons which begin at the exit window and travel toward the entrance window.  Note that all of the dependence on $\varphi$ is due to dot products of the $\vec a$ vectors.  Since these differ only in the sign of the $z$ component, they can all be rewritten in terms of squares of the components of $\vec a_1$, as in (\ref{e:Psi_sum}, \ref{e:Pabs_sum1}, \ref{e:Pabs_sum2}, \ref{e:Pabs_sum3}), and the $\varphi$ integral can be computed explicitly using
\begin{eqnarray}
\frac{1}{2\pi} \int_0^{2\pi} a_{1,x}^2 d\varphi
&=&
\cos^2(\theta) + \frac{3}{8} \sin^4(\theta)
\\
\frac{1}{2\pi} \int_0^{2\pi} a_{1,y}^2 d\varphi
&=&
\frac{1}{8} \sin^4(\theta)
\\
\frac{1}{2\pi} \int_0^{2\pi} a_{1,z}^2 d\varphi
&=&
\frac{1}{2} \sin^2(\theta) \cos^2(\theta).
\end{eqnarray}
We need only integrate numerically over one variable, $\theta$, in order to compute the decay and afterglow rates per particle.

We note that the magnetic field strength $B$ and the photon coupling $\bgam$ only appear in $\Gdec$ and $\Gaft$ through factors of $C^2$ in $|\Agam|^2$ and $\Pabs$.  This $C^2$ can be brought outside the integrals in (\ref{e:Gdec}) and (\ref{e:Gaft}).  Thus the decay and afterglow rates scale as $\Gdec, \Gaft \propto B^2 \bgam^2$, for any $\meff$, $\xiref$, and chamber geometry.  We need only compute the decay and afterglow rates for one value of each of $B$ and $\bgam$.

Next, we compute the expected flux of afterglow photons through the exit window for this simple afterglow experiment.  We assume that chameleon-photon oscillation is the dominant contributor to the total chameleon decay rate, and that all other decays are negligible.  The experiment proceeds in two stages: production and afterglow.  During the production stage, photons stream through the chamber at a rate of $\fgam$.  Each photon has a probability $\Pcg = C^2 \sin^2(\meff^2 L /(4k))$ of producing a chameleon particle, for a total production rate of $\fgam \Pcg$.  Meanwhile, if $N_\phi$ chameleons are present in the chamber, then the total decay rate is $N_\phi \Gdec$.  Thus the number of chameleons is given by $dN_\phi(t)/dt = \fgam \Pcg - N_\phi(t) \Gdec$.  Assuming that chameleon production occurs over a time interval $-\tp < t < 0$, we have 
\begin{equation}
N_\phi^\mathrm{(prod)}(t) 
=
\frac{\fgam \Pcg}{\Gdec} \left(1 - e^{-\Gdec(t+\tp)}\right) 
\end{equation}
in the production stage.  In the afterglow stage, $t>0$, the photon rate is reduced to zero, and the number of chameleons is 
\begin{equation}
N_\phi^\mathrm{(aft)}(t)
=
\frac{\fgam \Pcg}{\Gdec} \left(1 - e^{-\Gdec\tp}\right) e^{-\Gdec t}. 
\label{e:Naft}
\end{equation}
The expected signal during the afterglow stage of the experiment is found by multiplying $N_\phi$ by $\Gaft$,
\begin{equation}
\faft(t) 
=
\frac{\fgam \Pcg \Gaft}{\Gdec} \left(1 - e^{-\Gdec\tp}\right) e^{-\Gdec t}. 
\label{e:faft}
\end{equation}

\section{GammeV and GammeV--CHASE experiments}
\label{sec:gammev}

\subsection{GammeV geometry}
\label{subsec:gammev}

\begin{figure}[tb]
\begin{center}
\includegraphics[width=3.5in]{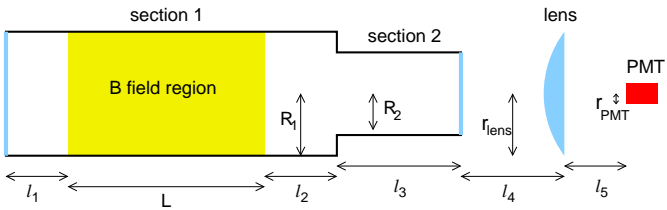}
\caption{Diagram (side view) of the apparatus used by GammeV and GammeV--CHASE to search for a chameleon afterglow.  Diagram is not to scale. Actual values of these lengths for each experiment are given in Table~\ref{t:apparatus}. \label{f:apparatus}}
\end{center}
\end{figure}

\begin{table}[tb]
\begin{center}
\begin{tabular}{|c|c|c|}
\hline
$\quad$quantity$\quad$ & $\quad$GammeV$\quad$ & $\quad$GammeV--CHASE$\quad$ \\
\hline
$\ell_1$               & $2.36$~m               & $2.0$~m       \\
$L$                    & $6.0$~m                & $6.0$~m       \\
$\ell_2$               & $1.16$~m               & $1.16$~m      \\
$\ell_3$               & $2.51$~m               & $0.30$~m      \\
$\ell_4$               & $2.03$~m               & $0.20$~m      \\
$\ell_5$               & $0.10$~m               & $0.10$~m      \\
$R_1$                  & $2.38$~cm              & $3.175$~cm    \\
$R_2$                  & $1.75$~cm              & $3.175$~cm    \\
$r_\mathrm{lens}$      & $2.54$~cm              & $2.54$~cm     \\
$r_\mathrm{PMT}$       & $2.5$~mm               & $2.5$~mm      \\
$t_0$                  & $1006$~sec             & $1$~sec       \\
$\Delta t$             & $3616$~sec             & $\sim 1000$~sec \\
$\fref$                & $0.53$                 & $0.53$        \\
$k\approx \omega$      & $2.33$~eV              & $2.33$~eV     \\
$B$                    & $5.0$~Tesla            & $5.0$~Tesla   \\
$V_\mathrm{pump}$      & $0.026$~m$^3$          & not available \\         
\hline
\end{tabular}
\caption{Properties of the GammeV and GammeV--CHASE experiments, including the dimensions shown in Fig.~\ref{f:apparatus}.  \label{t:apparatus}}
\end{center}
\end{table}

The actual apparatus used by GammeV to search for chameleons differs in a few ways from the simple experiment described above \cite{Chou_etal_2009}:
\begin{enumerate}
\item
the GammeV chamber has a second section, with a smaller radius, extending from $z=\ell_1+\ell_2+L$ to $z = \ell_1+\ell_2+L+\ell_3$, with the exit window at the end of the second section; 
\item
outside of the exit window is a lens that focuses the afterglow emerging from the chamber;
\item
afterglow photons must enter the aperture of the PMT, with radius $r_\mathrm{PMT}$, in order to be detected;
\item the PMT is uncovered a time $t_0$ after the laser has been turned off, and data are collected for a total time $\Delta t$;
\item the vacuum inside the chamber is maintained by a turbomolecular pump connected to a roughing pump, which increases the total volume accessible to chameleons by $V_\mathrm{pump}$;
\item chameleons light enough to enter the roughing pump will be removed from the chamber, so the range of masses which can be constrained is limited.
\end{enumerate}
Figure~\ref{f:apparatus} is a diagram of the GammeV apparatus, showing the lengths and radii defined above; the numerical values of these quantities are listed in Table~\ref{t:apparatus}.  We define $\ltotone= \ell_1+\ell_2+L$ to be the length of the first section, and $\ltottwo = \ltotone+\ell_3$ to be the total length of the chamber.

Limitations imposed on GammeV by the pumping system were discussed in \cite{Chou_etal_2009}.  A chameleon particle will be pumped out of the vacuum chamber if its effective mass $\meff(\textrm{rough})$ at the intake of the roughing pump, where the air pressure is $P_\mathrm{rough}=1.9\times 10^{-3}$~torr, is less than its energy $\omega \approx k$ inside the chamber.  This pumping occurs on time scales much smaller than $t_0$, so a chameleon model whose particles can be pumped out of the chamber is inaccessible to GammeV.  In particular, we shall see in Sec.~\ref{sec:constraints} that GammeV can only constrain a small subset of chameleon dark energy models.  For the remainder of this section we assume that $\meff(\textrm{rough}) > k$.  We proceed to calculate the decay and afterglow rates as functions of $\meff(\textrm{chamber})$, $\bgam$, and $\xiref$, noting that any resulting constraints on chameleon models will only be applicable when the above mass condition is met.

When calculating the afterglow rate, we must check whether the particle enters the second section, reaches the lens, and enters the aperture of the PMT. It is apparent that a particle beginning at the center of the entrance window, our chosen initial condition, will have a higher probability of being near the central axis of the chamber, and therefore a higher probability of reaching the second section, the lens, and the PMT.  Computing these probabilities for the GammeV geometry is straightforward; a particle beginning at an arbitrary position on the entrance window has a probability of $p_\mathrm{avg} = 2.04\times 10^{-5}$ of reaching the PMT, while a particle beginning at the center of the window has a probability of $p_\mathrm{ctr}=3.35\times 10^{-5}$.  Thus we must normalize our afterglow results by $\fgeom = p_\mathrm{avg} / p_\mathrm{ctr}$, which takes the value $0.610$ in this case.  Henceforth, we assume that $\Gaft$ has been normalized appropriately.  

\begin{figure}[tb]
\begin{center}
\includegraphics[angle=270,width=3.3in]{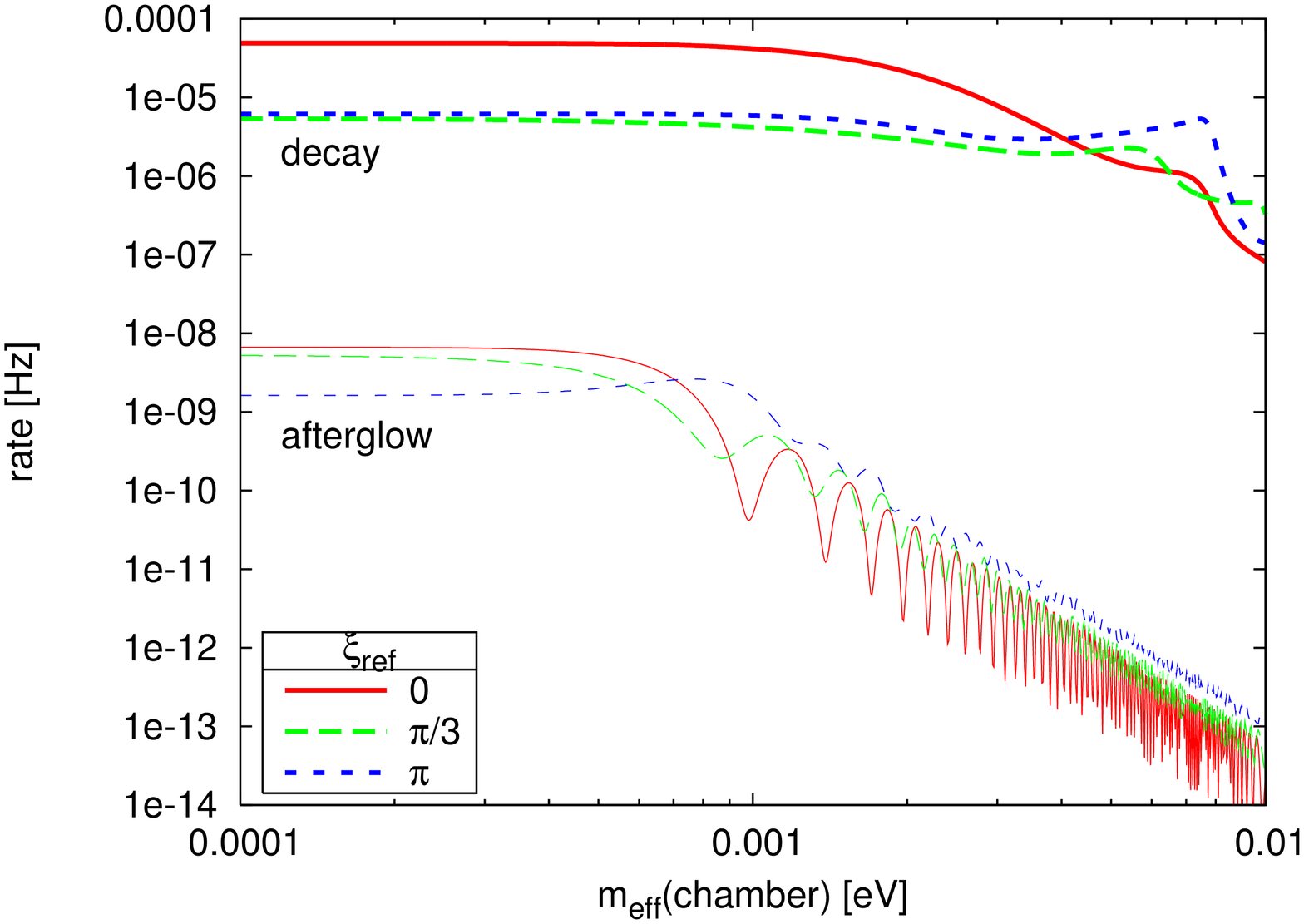}
\caption{$\Gdec$ (thick lines) and $\Gaft$ (thin lines) vs. $\meff$, for several values of $\xiref$, assuming a magnetic field of $B = 5$~Tesla and a photon coupling of $\bgam = 10^{12}$.\label{f:vary_mass}}
\end{center}
\end{figure}

\begin{figure}[tb]
\begin{center}
\includegraphics[angle=270,width=3.3in]{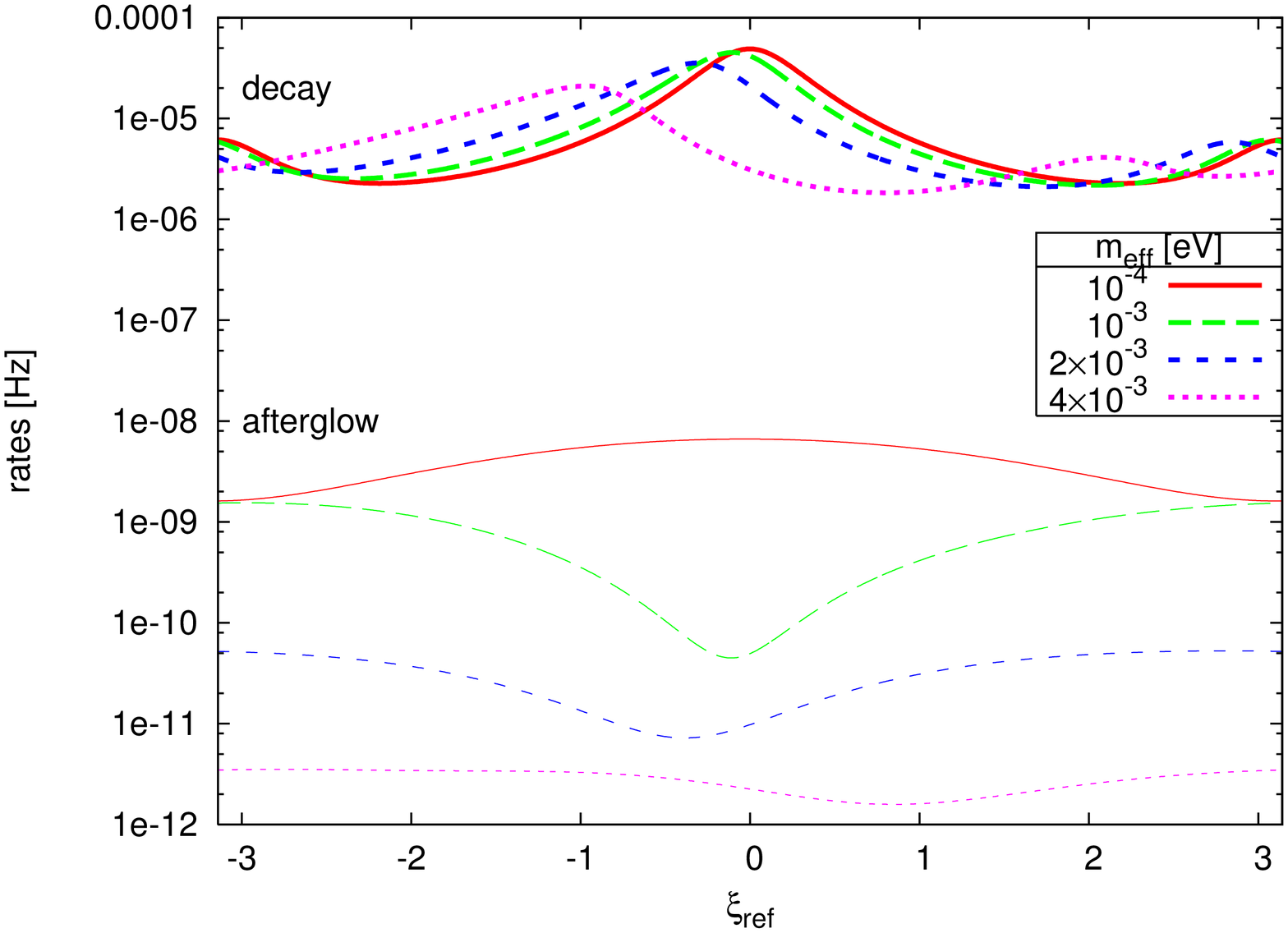}
\caption{$\Gdec$ (thick lines) and $\Gaft$  (thin lines) vs. $\xiref$, for several values of $\meff$, assuming a magnetic field of $B = 5$~Tesla and a photon coupling of $\bgam = 10^{12}$.\label{f:vary_xiref}}
\end{center}
\end{figure}

The final results of the computation detailed above are the afterglow rate $\Gaft$ per chameleon particle, and the rate $\Gdec$ of the decay of a chameleon particle to a photon,  both of which are functions of the background magnetic field $\vec B = B\xhat$, the chameleon mass $\meff(\textrm{chamber})$, the chameleon-photon coupling $\bgam$, and the phase shift $\xiref$ at each wall reflection.  As noted earlier, $\Gdec$ and $\Gaft$ both scale as the square of the magnetic field strength and the photon coupling.  The variations of these rates with $\meff(\textrm{chamber})$ and $\xiref$ are more complicated.  

Figure~\ref{f:vary_mass} shows $\Gdec$ and $\Gaft$ as functions of mass for several values of $\xiref$. Consider first the thin, solid line, which shows $\Gaft$ for $\xiref=0$.  The afterglow rate is dominated by chameleons on trajectories with small $\theta$, which bounce no more than a few times from the chamber walls, since these are the most likely to produce photons which reach the detector.  For a trajectory with no bounces, (\ref{e:Pcg}) implies that the afterglow rate is proportional to $\sin^2(\meff^2 L / (4k))$.  Thus we expect $\Gaft$ to be small when $\meff = \sqrt{4 \pi j k / L} = 9.81 \times 10^{-4} j^{1/2}$~eV for any positive integer $j$; at these masses, destructive interference in the chameleon-photon oscillation suppresses photon production.  As expected, Fig.~\ref{f:vary_mass} shows the first local minimum of the afterglow rate at $\meff\approx 0.001$~eV and the fourth around $0.002$~eV.  Furthermore, these local minima are very sharp, since paths with slightly larger $\theta$ differ only slightly in phase at the end of the magnetic field region.  In contrast, the other two afterglow rate plots have much shallower minima at different masses.  This is because a nonzero $\xiref$ implies that paths with different numbers of bounces can exit the magnetic field region with very different phases.  Meanwhile, the decay rates, plotted as thick lines in Fig.~\ref{f:vary_mass}, have fewer features, since they average paths over a much larger range of angles.

$\Gdec$ and $\Gaft$ are shown as functions of $\xiref$, for a few fixed masses, in Fig.~\ref{f:vary_xiref}.  In the low-mass limit, the only phase shift between the chameleon and the photon is due to reflection from the walls.  A total phase difference of zero implies maximal constructive interference between the chameleon and photon wavefunctions, and, hence, maximal chameleon-photon oscillation. Thus the decay and afterglow rates will peak at $\xiref=0$.  This is consistent with the plots of $\Gdec$ and $\Gaft$ for $\meff = 10^{-4}$~eV, shown as thick and thin solid lines, respectively.  At larger masses, this maximum in the decay rate shifts to negative $\xiref$, in order to compensate for the positive phases $\eL$, $\xiprop$, and $\eR$ caused by a nonzero chameleon-photon mass difference.  Meanwhile, at $\meff = 0.001$~eV, $\xiref\approx 0$ corresponds to maximal destructive interference, as discussed above.  Thus a nonzero $\xiref$ increases $\Gaft$ by making the interference in chameleon-photon oscillation more constructive, leading to a minimum around $\xiref=0$, as shown in Fig.~\ref{f:vary_xiref}.  

Next, we will show that the effects of the initial conditions on the dynamics of chameleon-photon oscillation are negligible; the normalization $\fgeom$ is the only correction that we need to make to the afterglow rate.  Since $z=0$ and $|\vec k|$ are fixed, the full set of possible initial conditions on the entrance window is described by four parameters: $x$, $y$, $\theta$, and $\varphi$.  Until now, we have restricted ourselves to a two-parameter subset, $x=y=0$.  Our subset consists of paths that remain in the same plane as they bounce around, a plane which contains the $z$ axis.  This condition will remain true if we allow nonzero initial $x$ and $y$, subject to the constraint $y/x = \tan(\varphi)$.  Thus we can study a third parameter out of the four.  This can be done through a simple modification of our previous calculation.  Assume that the chamber were extended by a length $\ell_0$ in the negative $z$ direction, with the entrance window now at $z=-\ell_0$.  A particle beginning at the center of this new entrance window would, when it reached $z=0$, satisfy $y/x = \tan(\varphi)$, with $x$ and $y$ not necessarily zero.  If we compute $\Pabs$ and $|\Agam(\ell_B)|$ using these new paths, while using the chamber length without $\ell_0$ to find the time taken by each path, then we obtain decay and afterglow rates that can be compared with our previous calculations.  Table~\ref{t:l0} shows that varying $\ell_0$, from zero to much larger than the length of the entire chamber, leaves the decay rate essentially unchanged, and changes the afterglow rate by no more than a few percent.

\begin{table}[tb]
\begin{center}
\begin{tabular}{|c|c|c|}
\hline
$\quad\ell_0$ [m]$\quad$ & $\qquad\Gdec$ [Hz]$\qquad$ & $\qquad\Gaft$ [Hz]$\qquad$ \\
\hline
0            & $4.905\times 10^{-5}$    & $6.643\times 10^{-9}$ \\
1            & $4.899\times 10^{-5}$    & $6.865\times 10^{-9}$ \\
10           & $4.903\times 10^{-5}$    & $6.845\times 10^{-9}$ \\
100          & $4.897\times 10^{-5}$    & $6.876\times 10^{-9}$ \\
\hline
\end{tabular}
\caption{Decay and afterglow rate vs. $\ell_0$.  These have been computed for the GammeV geometry, assuming a reflectivity of $\fref=0.53$, a magnetic field of $B=5$~Tesla, a chameleon mass in the chamber of $\meff=10^{-4}$~eV, and a chameleon-photon coupling of $\bgam=10^{12}$. \label{t:l0}}
\end{center}
\end{table}

\begin{figure}[tb]
\begin{center}
\includegraphics[width=2.5in]{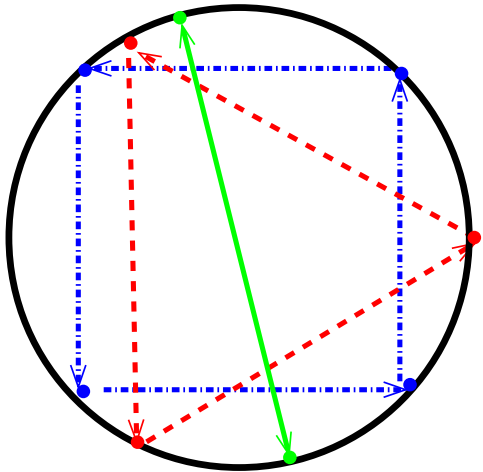}
\caption{Projection into the $xy$ plane of a 2-point path (solid green line), a 3-point path (dashed red line) and a 4-point path (dash-dotted blue line).  \label{f:n-point}}
\end{center}
\end{figure}

Each path in the three-parameter subset considered above was confined to a single plane.  Other paths do not remain in a plane, but travel through the chamber in a corkscrew fashion.  It is useful to consider the projection of these paths onto the $xy$ plane, as shown in Fig.~\ref{f:n-point}.  Let us also assume for this discussion that $\theta$ is not small, so that each path bounces multiple times.  The paths confined to a plane would, in this projection, appear to bounce back and forth between the same two points, passing through the center between any two bounces, as shown by the solid line in Fig.~\ref{f:n-point}.  We call such paths 2-point paths.  If the condition $y/x = \tan(\varphi)$ were violated by a small amount, then the points between which the path bounced would appear to move slightly as the particle progressed down the chamber; the solid line in Fig.~\ref{f:n-point} would appear to precess.  For larger deviations, the $xy$ projection of the path would close on itself, forming an equilateral triangle, as shown by the dashed line in Fig.~\ref{f:n-point}; we call this a 3-point path.  We can have $n$-point paths, for any $n \geq 2$, which trace out equilateral polygons of $n$ vertices.  Such paths are interesting because they allow us to carry out explicitly the sums in (\ref{e:psi_Bexit}) and (\ref{e:Pabs}); for an $n$-point path, each of the $\vec a$ vectors is equal to one of the first $n$ vectors.

$n$-point paths for $n\geq 3$ are different from 2-point paths in that they avoid the center; 3-point paths remain at a distance of at least $R_1/2$ from the central axis, and $n$-point paths for $n>3$ remain even farther from the center.  Another difference is that paths with higher $n$ bounce more frequently for the same $\theta$, since they travel a shorter distance between bounces.  

In order to test the effects of these differences on our rate calculations, we compute 3-point decay and afterglow rates by summing (\ref{e:psi_Bexit}) and (\ref{e:Pabs}) for 3-point paths.  Our assumption that the particle begins at the center of the entrance window gives a valid approximation to the decay and afterglow rates only if $\Gdec(\textrm{2-point})\approx \Gdec(\textrm{3-point})$ and $\Gaft(\textrm{2-point})\approx \Gaft(\textrm{3-point})$.  Our assumption will be conservative if $\Gdec(\textrm{2-point}) > \Gdec(\textrm{3-point})$ and $\Gaft(\textrm{2-point}) < \Gaft(\textrm{3-point})$.

We find that neither the avoidance of the center nor the shorter distance between bounces affects the afterglow rate very much.  The fact that $n$-point paths for $n>2$ avoid the center is accounted for in the normalization factor $\fgeom$ for these paths.  Also, $\Gaft$ is dominated by paths with small $\theta$, which are more likely to reach the PMT.  Such paths bounce only a small number of times, and the computed $\Gaft$ does not change appreciably.  The 3-point afterglow rate is $\Gaft(\textrm{3-point}) = 7.005\times 10^{-9}$~Hz for the parameters used in Table~\ref{t:l0}.   This is within a few percent of $\Gaft(\textrm{2-point}) = 6.643\times 10^{-9}$~Hz.

We also find that the decay rate is somewhat lower for 3-point paths, assuming low chameleon masses.  For the GammeV geometry and the parameters $\fref=0.53$, $B=5$~Tesla, $\meff(\textrm{chamber})=10^{-4}$~eV, and $\bgam = 10^{12}$, we find $\Gdec(\textrm{3-point})= 3.800\times 10^{-5}$~Hz, compared to $\Gdec(\textrm{2-point})=4.905\times 10^{-5}$~Hz.  This is due to the fact that 3-point paths bounce more frequently, suppressing the coherent buildup of photon amplitude over multiple bounces.  In our constraints, we use the 2-point computation of the decay rate.  Since an overestimate of $\Gdec$ leads to an underestimate of the expected signal $N_\phi(t)\Gaft$, this approximation is slightly conservative.

At higher masses, the 3-point paths make a greater contribution to the decay rate.  As $\meff$ is increased beyond $\sqrt{4\pi k / L} \approx 10^{-3}$~eV, destructive interference suppresses oscillation in small-$\theta$ paths.  Large-$\theta$ paths, which travel shorter distances between bounces, become more important to the computation of $\Gdec$.  The shortest possible distance between bounces for a 2-point path is $2R_1$, which occurs when $\theta = \pi/2$.  For $\meff \sim \sqrt{4\pi k / (2R_1)} \approx 10^{-2}$~eV, oscillation is suppressed by destructive interference even for these paths.  For a 3-point path, however, the shortest distance between bounces is only $R_1\sqrt{3}$.  Thus $\Gdec(\textrm{3-point})$ will exceed $\Gdec(\textrm{2-point})$ at some mass $\meff \lesssim 10^{-2}$~eV.  Beyond this mass, our approximation of the decay rate by $\Gdec(\textrm{2-point})$ is no longer conservative.  

GammeV, which probed masses up to $10^{-3}$~eV, remained well within the realm of validity of this approximation; $\Gdec \leq \Gdec(\textrm{2-point})$ for $\meff < 10^{-3}$~eV.   Thus $ \Gdec(\textrm{2-point})$ was used to approximate $\Gdec$ in the analysis of ref.~\cite{Chou_etal_2009} in order to provide conservative bounds on the chameleon parameter space.  We will revisit the issue in Sec.~\ref{subsec:gammev-chase}, where we shall see  that this approximation breaks down at the largest masses probed by GammeV--CHASE.

\subsection{GammeV--CHASE}
\label{subsec:gammev-chase}

As discussed in \cite{Chou_etal_2009,Steffen_Upadhye_2009}, the original GammeV chameleon search was constrained by four different technical limitations:
\begin{enumerate}
\item destructive interference in the $L=6$~m magnetic field length suppressed the production of chameleons with masses greater than $\sqrt{4\pi\omega/L} \approx 10^{-3}$~eV;
\item systematic uncertainties in the PMT dominated the total error, weakening constraints at low $\bgam$;
\item the transition between filling the cavity and collecting afterglow data required $t_0 = 1006$~sec after the laser was turned off, diminishing sensitivity to high $\bgam$ (rapidly decaying chameleons);
\item the roughing pump in the vacuum system exhausted to the room, which meant that the lowest-density ``wall'' of the chamber was the $P=1.9\times 10^{-3}$~torr intake of the roughing pump.
\end{enumerate}
GammeV--CHASE improves considerably upon GammeV by addressing each one of these limitations.

First, glass windows will divide the magnetic field region into partitions of different lengths, $0.3$~m, $1.0$~m, and $4.7$~m.  These three partitions remove regions of insensitivity since the chameleon-photon oscillation lengths for each partition are not commensurate with each other for many multiples.  Also, since the partitions are somewhat shorter than the original 6m cavity, they provide some sensitivity to larger mass chameleons, up to a few meV.  This improvement is especially significant because chameleon masses at the dark energy scale $2.4 \times 10^{-3}$~eV, were inaccessible to the previous experiment.

Improvements to the optical system will allow GammeV--CHASE to push to both lower and higher $\bgam$.  Modulating the PMT signal using a mechanical shutter allows the detector noise to be monitored.  Since systematic uncertainty in the PMT dark rate was the dominant source of noise in GammeV, the sensitivity to low afterglow rates will improve by roughly an order of magnitude in GammeV--CHASE. 
Meanwhile, sensitivity to high $\bgam$ will improve by three
orders of magnitude with the incorporation of two changes.  First, 
a more rapid transition will be made between filling the cavity and collecting data, shortening the dead time by a factor of more than 100 and improving sensitivity to larger $\bgam$ by almost two
orders of magnitude.  Second, data will be collected at lower magnetic
fields of $1$ and $0.2$~Tesla, in addition to the original field of $5$~Tesla.  The low magnetic field provides an additional
order of magnitude in sensitivity because it slows conversion of chameleon particles to photons; this maintains a detectable population of chameleons while the transition to data acquisition occurs.

Finally, improvements to the pumping system will take two forms.  First, the vacuum pressure will be reduced by approximately three orders of magnitude from $\sim 10^{-7}$ torr to $\sim 10^{-10}$ torr.  Second, the vacuum system will not exhaust to the room as it did in the original experiment.  These improvements come through the use of ion pumps placed at strategic locations on the apparatus, as well as cryogenic pumping within the magnetic field region; residual gases will freeze to the bore of the magnet.  The fact that this system does not exhaust to the room will mean that chameleons need only bounce from the chamber walls, with densities $\rho \sim 1$~g/cm$^3$, rather than the much stronger condition that they bounce before the intake of the roughing pump, with $\rho \sim 10^{-9}$~g/cm$^3$.  Improvements to the pumping system will allow GammeV--CHASE to probe chameleons whose masses scale much more slowly with density.  For chameleons with $\meff \propto \rho^\eta$, GammeV--CHASE will probe $\eta \gtrsim 0.2$. Moreover, the improvement in constraints will be qualitative as well as quantitative.  The large range of masses probed will mean that constraints on the photon coupling will be only weakly dependent on the matter coupling, and vice versa, as we shall see.

\begin{figure}[tb]
\includegraphics[width=3.2in]{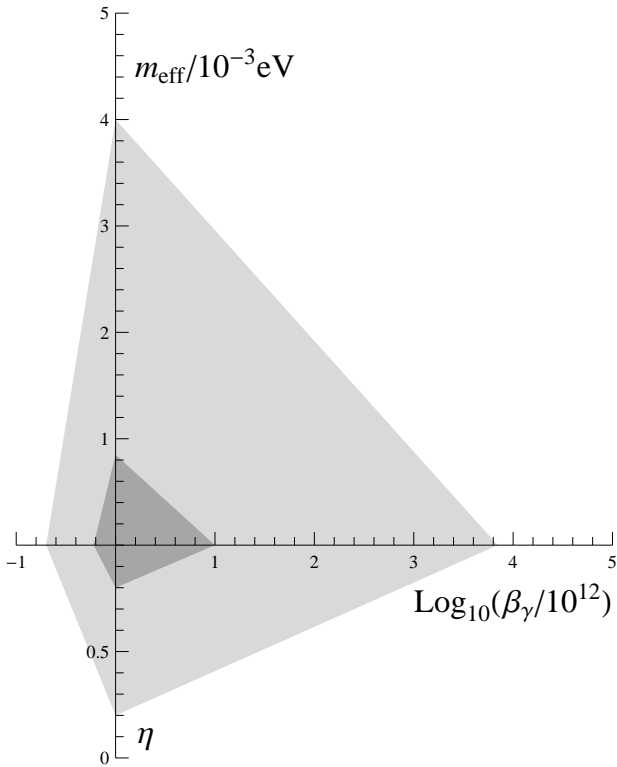}
\caption{Improvements in sensitivity to $\bgam$, $\eta$, and $m_{\text{eff}}$ with GammeV--CHASE.  Low $\bgam$ improves by roughly an order of magnitude.  High $\bgam$ improves by more than three orders of magnitude.  The parameter $\eta$ improves from 0.8 to nearly 0.2.  The mass sensitivity improves by a factor of nearly four. \protect\label{f:extensions}}
\end{figure}

To summarize, there are essentially four important directions in which the sensitivity of GammeV--CHASE improves upon GammeV: smaller $\bgam$, larger $\bgam$, smaller $\eta$, and larger $\meff(\textrm{chamber})$.  Figure~\ref{f:extensions} is a "radar chart" of the improvements in these parameters between the two experiments.

We showed in Sec.~\ref{subsec:gammev} that our choice of initial conditions, chameleon particles which begin at the center of the entrance window, resulted in approximations to $\Gdec$, $\Gaft$, and the total afterglow flux that were accurate, and slightly conservative, for GammeV.  However, at larger chameleon masses, we expect these approximations to break down as trajectories with smaller distances between bounces contribute more to the decay rate.  This is because destructive interference suppresses chameleon-photon oscillation on segments of the particle trajectory that are longer than $4\pi k / \meff^2$.  For example, in GammeV and GammeV--CHASE, this length corresponds to $23$~cm when $\meff = 5\times 10^{-3}$~eV.

The n-point trajectories discussed earlier, for $n \geq 3$, travel smaller distances between bounces.  For example, 3-point trajectories have a minimum distance between bounces of $R_1\sqrt{3}$, compared to $2R_1$ for the 2-point trajectories resulting from our initial conditions.  At greater $\meff$, $n$-point trajectories with greater $n$ will become important.  By comparing the 2-point and 3-point calculations of the decay rate, we estimate the mass at which the 2-point decay rate is no longer a conservative approximation to the total decay rate.  Beyond this mass, at which, $\Gdec(\textrm{3-point})=\Gdec(\textrm{2-point})$, our computation of the decay rate becomes increasingly inaccurate.

\begin{figure}[tb]
\begin{center}
\includegraphics[angle=270,width=3.3in]{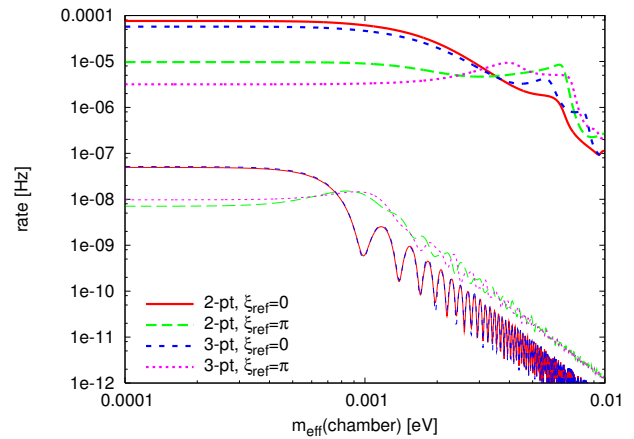}
\caption{$\Gdec$ (thick lines) and $\Gaft$ (thin lines) vs. $\meff$, for 2-point and 3-point paths with two different values of $\xiref$.  We assume $B = 5$~Tesla and $\bgam = 10^{12}$.  For simplicity, we use a geometry similar to that of GammeV--CHASE but with only one $L=6$~m partition in the magnetic field region. \label{f:n-point_rates}}
\end{center}
\end{figure}

Figure~\ref{f:n-point_rates} shows 2-point and 3-point computations of the decay and afterglow rates, for an afterglow experiment with a single $6$~meter partition in the magnetic field region.  At low mass, it is clear from the figure that our approximations $\Gdec \approx \Gdec(\textrm{2-point})$ and $\Gaft \approx \Gaft(\textrm{2-point})$ are excellent.  Where the 2-point and 3-point computations differ, our approximations are conservative; we slightly underestimate the afterglow rate and overestimate the decay rate.  At larger masses, the 2-point decay rates begin to drop, and the 3-point rates catch up.  $\Gdec(\textrm{3-point})$ exceeds $\Gdec(\textrm{2-point})$ at $\meff \approx 4\times 10^{-3}$~eV for $\xiref=0$, and at $\meff \approx 3\times 10^{-3}$~eV for $\xiref=\pi$; we find similar masses for other values of $\xiref$.

At larger masses, we must compute the decay and afterglow rates by averaging (\ref{e:Gdec}) and (\ref{e:Gaft}) over all initial positions ($x$, $y$) on the entrance window.  The symmetries that allowed us to sum explicitly the series in (\ref{e:psi_Bexit}) and (\ref{e:Pabs}) will no longer be present, making the integration more computationally intensive.  In this work, for the purposes of forecasting GammeV--CHASE constraints, we simply cut off our constraints at the mass where $\Gdec(\textrm{3-point})$ first exceeds $\Gdec(\textrm{2-point})$.


\section{Chameleon models: constraints and forecasts} 
\label{sec:constraints}

\subsection{Model-independent constraints}
\label{subsec:model-independent_constraints}

The decay rate (\ref{e:Gdec}) and the afterglow rate (\ref{e:Gaft}) depend on three properties of the chameleon particle: its photon coupling $\bgam$, its phase shift $\xiref$ at each wall reflection, and its effective mass $\meff(\textrm{chamber})$ inside the vacuum chamber.  Constraints on these parameters are model-independent in the sense that they do not depend on a knowledge of the chameleon potential $V(\phi)$.  

Also important is the requirement that chameleon particles be contained inside the vacuum chamber, which implies that the effective mass $\meff(\textrm{wall})$ inside the chamber walls is greater than the chameleon energy $\omega = 2.33$~eV.  If we assume that the scaling of effective mass with density can be approximated by a power law $\meff(\rho) \propto \rho^\eta$ over the densities of interest, then the smallest value of $\eta$ that can be constrained is
\begin{equation}
\eta_\mathrm{min}
=
\frac{\log\left[\meff^\mathrm{(min)}(\textrm{wall})\right] 
    - \log\left[\meff^\mathrm{(max)}(\textrm{chamber})\right]}
  {\log\left[\rho(\textrm{wall})\right] 
    - \log\left[\rho(\textrm{chamber})\right]}.
\label{e:eta_min}
\end{equation}
For $\eta > \eta_\mathrm{min}$, a larger range of chameleon masses can be excluded.

\subsubsection{GammeV}

GammeV \cite{Chou_etal_2009} constrained the afterglow signal (\ref{e:faft}) averaged over the observation window, which began at a time $t_0 = 1006$~sec after the laser was turned off, and had a duration $\Delta t = 3616$~sec.  A systematic uncertainty of $\sigma_\Gamma = 12.0$~Hz in the PMT dark rate $\Gamma_\mathrm{dark} = 115$~Hz meant that GammeV could rule out to $3\sigma$ any chameleon model with an average afterglow greater than $36$~Hz in the time window.  

\begin{figure}[tb]
\includegraphics[angle=270,width=3.3in]{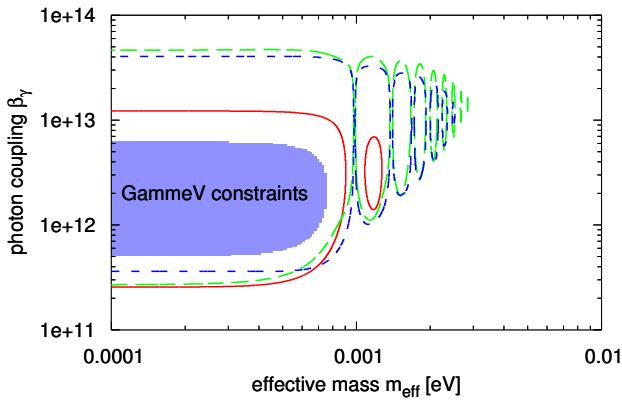}
\caption{Model-independent $3\sigma$ constraints from GammeV.  The shaded region shows the phase-independent constraints from \cite{Chou_etal_2009}, which computed the afterglow rate using only non-bouncing trajectories.  The solid (red), long dashed (green), and short dashed (blue) lines correspond to $\xiref = 0$, $\pi/3$, and $\pi$, respectively.  Note that these constraints apply only to chameleons which can be contained in the vacuum chamber, $\meff(\textrm{rough})>k$, as discussed in Sec.~\ref{sec:gammev}. \label{f:gammev1}}
\end{figure}

Figure \ref{f:gammev1} shows our model-independent constraints for GammeV.  The shaded region shows GammeV constraints from \cite{Chou_etal_2009}.  These are not only model-independent but also phase-independent; we considered only non-bouncing trajectories when evaluating $\Gaft$, in order to remove the effects of phase shifts, and assumed $\xiref=0$ when calculating $\Gdec$ in order to overestimate the decay rate.  Since a larger decay rate suppresses the expected signal $\faft$ found in (\ref{e:faft}), this set of assumptions is conservative, and applies to all $\xiref$.  

Meanwhile, the solid and dashed lines in Fig.~\ref{f:gammev1} show the regions of parameter space excluded when we consider bouncing trajectories in the calculation of $\Gaft$ and assume a specific value for $\xiref$.  Since the decay rate $\Gdec$ is dominated by paths which bounce hundreds times, while paths contributing to the afterglow rate $\Gaft$ bounce only $\sim 1$ times, $\Gdec$ depends much more strongly on $\xiref$ than does $\Gaft$.  Thus, at high $\bgam$, where our constraints are limited by rapid chameleon decays, the suppression of $\Gdec$ due to nonzero $\xiref$ extends our constraints to stronger photon couplings.  On the other hand, at low $\bgam$, the decay time $1/\Gdec$ is much longer than the duration of the experiment, making decays irrelevant.  In this regime, the slight suppression of $\Gaft$ at nonzero $\xiref$ means that the $\xiref=0$ constraints are the strongest.

The constrained regions contain several ``islands'' at $\meff \gtrsim 10^{-3}$~eV.  These are caused by the zeros of the chameleon production probability, given by (\ref{e:Pcg}) with $t=L$.  Photons of energy $k=2.33$~eV passing through a magnetic field region of length $L=6$~m cannot produce chameleons with masses $\meff = \sqrt{4 \pi j k / L}$, for any positive integer $j$, due to total destructive interference in photon-chameleon oscillation.

The range of $\eta$ values in GammeV was severely limited by the pumping system used to maintain the vacuum inside the chamber.  The lowest-density ``wall'' was the intake of the roughing pump, with $\rho_\mathrm{wall} = 3.0\times 10^{-9}$~g/cm${}^3$; chameleons too light to reflect from this intake would be pumped out of the chamber.  For a reflection phase of $\xiref = 0$, the largest mass probed by GammeV was $\meff^\mathrm{(max)}(\textrm{chamber}) = 1.27\times 10^{-3}$~eV, giving $\eta_\mathrm{min} = 0.76$.  

\subsubsection{GammeV--CHASE}
\label{subsubsec:gammev-chase}

GammeV--CHASE will monitor the PMT dark rate in real time by modulating the afterglow signal from the chamber.  Furthermore, rapid switching of the PMT will allow the experiment to begin data collection at $t_0 \approx 1$~sec, so that GammeV--CHASE will probe chameleon theories with much higher decay rates than those excluded by GammeV.  Here, we estimate the constraints that will be obtained by GammeV--CHASE by averaging the signal over a time window of $\Delta t = 20$~sec, assuming that the PMT is modulated with a duty cycle of $0.5$.  The uncertainty in the dark rate $\Gamma_\mathrm{dark} \approx 100$~Hz over this interval will be $\sigma_\Gamma \approx 3.16$~Hz.

\begin{figure}[tb]
\includegraphics[angle=270,width=3.3in]{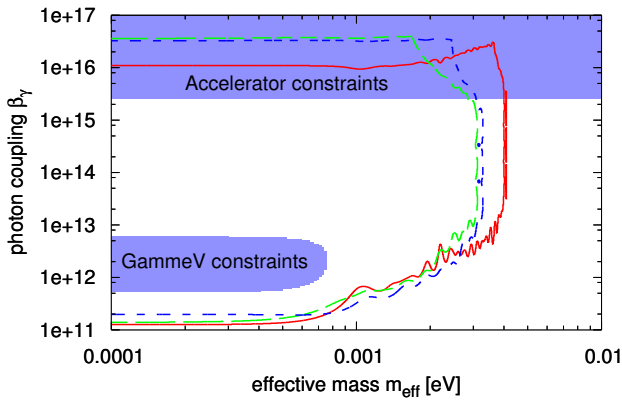}
\caption{Forecast model-independent $3\sigma$ constraints from GammeV--CHASE.  The shaded regions show current constraints from GammeV \cite{Chou_etal_2009} as well as the accelerator constraints of \cite{Brax_etal_2009}.  The solid (red), long dashed (green), and short dashed (blue) lines correspond to $\xiref = 0$, $\pi/3$, and $\pi$, respectively. \label{f:gammev-chase}}
\end{figure}

Forecast constraints from GammeV--CHASE are shown in Figure~\ref{f:gammev-chase}, for $\xiref=0$, $\pi/3$, and $\pi$.  The shaded regions show current constraints from GammeV \cite{Chou_etal_2009} and particle accelerators \cite{Brax_etal_2009}.  GammeV--CHASE is expected to bridge the gap between GammeV constraints and accelerator constraints for masses around the dark energy scale, $\meff \sim 10^{-3}$~eV.

The constraints shown are limited at low $\bgam$ by the sensitivity to low, nearly constant chameleon fluxes.  Consider, for example, $\xiref=0$.  From Fig.~\ref{f:vary_mass} and the proportionality $\Gdec$,~$\Gaft \propto \bgam^2$, we see that $\Gdec \sim 10^{-6}$~Hz and $\Gaft \sim 10^{-10}$~Hz when $\bgam = 10^{11}$ and $\meff$ is small.  The corresponding decay time $\Gdec^{-1}$ is much larger than any time scale associated with the experiment, meaning that the number of chameleons in the chamber, given by (\ref{e:Naft}), remains nearly constant at $N_\phi^\mathrm{(aft)} \approx \fgam \Pcg \tp \sim 10^{11} (\bgam/10^{11})^2$ for the duration of the experiment.  At low $\bgam$ and $\meff$, the total afterglow flux reaching the detector is $N_\phi^\mathrm{(aft)}\Gaft \sim 10 (\bgam/10^{11})^4$~Hz, which drops below our sensitivity $3\sigma_\Gamma \approx 9.5$~Hz around $\bgam = 10^{11}$.  Meanwhile, at high $\bgam$, the constraints are limited by rapid decays, $\Gdec^{-1} \ll t_0 = 1$~sec, which reduce the chameleon population before our detector can be switched on.  Again assuming $\xiref=0$ and low $\meff$, we see from Fig.~\ref{f:vary_mass} that $\Gdec \sim 10^{-4} (B/5\textrm{T})^2 (\bgam/10^{12})^2$~Hz.  At the lowest magnetic field used, $B = 0.2$~Tesla, the corresponding decay time $\Gdec^{-1}$ drops below $t_0$ for $\bgam \sim 10^{16}$, the upper limit of our constrained region.  Finally, as with GammeV, we see that the $\xiref=0$ constraints are strongest at low $\bgam$, while constraints at larger phase shifts extend to higher $\bgam$.  

At high $\meff$, our constraints are cut off by uncertainties in the calculation of the decay rate.  Our approximation that $\Gdec < \Gdec(\textrm{2-pt})$ breaks down around $\meff \approx 4 \times 10^{-3}$~eV, where $\Gdec(\textrm{3-pt})$ exceeds $\Gdec(\textrm{2-pt})$.  Had we instead approximated $\Gdec \approx \max(\Gdec(\textrm{2-pt}), \Gdec(\textrm{3-pt}))$, our constraints would have extended to $6 \times 10^{-3}$~eV.

Compared with the GammeV constraint plot in Fig.~\ref{f:gammev1}, the GammeV--CHASE forecast constraints in Fig.~\ref{f:gammev-chase} have few sharp features such as islands.  This is due to the use of partitions in the magnetic field region, as well as data runs at multiple magnetic field values.  The partitions ensure that there are no zeros in the total chameleon production probability (\ref{e:Pcg}) in the range of masses probed.  A chameleon whose mass prevents it from being produced in the $4.7$~m partition, for example, may still be produced in the $1.0$~m partition or the $30$~cm partition.  Multiple magnetic field values ensure overlap between constraints from the different runs.  For example, the greatest $\bgam$ probed by the $B = 5$~Tesla run will be larger than the smallest $\bgam$ probed by the $B=1$~Tesla run, ensuring a continuous constrained region.  Overlaps between multiple magnetic fields and multiple partitions allow GammeV--CHASE to smooth over the features that are present in the decay rate, and especially in the afterglow rate, as in Fig.~\ref{f:vary_mass}.

Modifications to the vacuum system, discussed earlier in Sec.~\ref{subsec:gammev-chase}, will drastically improve the range of $\eta$ values probed.  The lowest-density wall will now consist of the glass windows, with $\rho_\mathrm{wall} \approx 1$~g/cm${}^3$.  Inside the magnetic field region, the operating temperature of $4$~K and pressure of $10^{-10}$~torr will reduce the matter density inside the chamber to $\rho_\mathrm{vac} = 8 \times 10^{-16}$~g/cm${}^3$.  For chameleons with $\bgam \ll \bmat$, we can neglect the energy density associated with the magnetic field, and  approximate $\meff \propto \rhom^\eta$.  In that limit, $\eta_\mathrm{min} = 0.18$.  For $\bgam \sim \bmat$, the magnetic field becomes important, since $\rhog = 1.11\times 10^{-13}$~g/cm${}^3$ is greater than the matter density.  In this case, a mass scaling $\meff \propto (\rhom + \rhog)^\eta$ is more appropriate, and we find $\eta_\mathrm{min} = 0.21$.

\subsection{Power law chameleons}
\label{subsec:power_law_chameleons}

Consider a chameleon theory with potential $V(\phi) = g \phi^{\mathcal N}$, for $g>0$ and ${\mathcal N}>2$.  Since $\bgam\phi/\Mpl, \bmat\phi/\Mpl \ll 1$, in order for the theory to be consistent with fifth force constraints, the effective potential is given by
\begin{equation}
\Veff(\phi) = g\phi^{\mathcal N} - \frac{\phi}{\Mpl}(\bmat\rhom + \bgam\rhog).
\end{equation}
In a bulk medium of constant $\rhom$ and $\rhog$, the chameleon mass is evaluated at the minimum of this potential,
\begin{eqnarray}
\phimin
&=&
\left(\frac{\bmat\rhom + \bgam\rhog}{{\mathcal N}g\Mpl}\right)^\frac{1}{{\mathcal N}-1}
\label{e:phimin_pwrlaw}
\\
\meff
&=&
\sqrt{g {\mathcal N} ({\mathcal N}-1)} 
\left(\frac{\bmat\rhom + \bgam\rhog}{{\mathcal N} g \Mpl}\right)^\frac{{\mathcal N}-2}{2{\mathcal N}-2}.
\label{e:meff_pwrlaw}
\end{eqnarray}
Note that the mass scales with density as $\meff \propto (\bmat\rhom + \bgam\rhog)^\eta$ with $\eta = \frac{{\mathcal N}-2}{2{\mathcal N}-2}$.  For $\phi^4$ theory we have $\eta = 1/3$.  Although $\eta$ grows with ${\mathcal N}$, it asymptotically approaches $1/2$ at large ${\mathcal N}$.  Thus, the original GammeV experiment could not exclude any of these power law models.  GammeV--CHASE, on the other hand, will impose strong constraints on power law chameleons, as we will show for the $\phi^4$ model.

Ref.~\cite{Brax_etal_2007} finds that the phase shift for power law chameleon models, for any real ${\mathcal N}$, is given by
\begin{equation}
\xiref 
=
\frac{\pi}{2}
\left( \left|\frac{3{\mathcal N}-2}{{\mathcal N}-2}\right| - 1 \right).
\label{e:xiref}
\end{equation}
In particular, for $\phi^4$ theory we find $\xiref = 2\pi$, equivalent to zero.  The inverse power law potential $V \propto 1/\phi$ has a phase of $\xiref = \pi/3$.  In the limit of large $|{\mathcal N}|$, $\xiref \rightarrow \pi$.  We also find that the exponential potential $V \propto \exp(\phi/M)$ has $\xiref = \pi$.

Finally, for a power law chameleon with ${\mathcal N} \geq 3$ a real number, we must consider the possibility of chameleon fragmentation, in which two chameleon particles can interact to produce more than two chameleon particles.  The PMT is sensitive to photons with energy $\omega \approx 2.33$~eV, the energy of our laser.  Repeated chameleon fragmentation results in a population of low energy chameleons, whose afterglow photons are too low in energy to be observed by the PMT.  Assuming a fragmentation cross section $\sfrag$, the fragmentation rate is $\Gfrag = N_\phi \sfrag / V$, where $V = 0.030$~m${}^3$ is the volume of the chamber.  Here, $N_\phi$ represents the number of chameleons energetic enough that their afterglow photons will be observable by the PMT; although fragmentation increases the total number of chameleon particles, it decreases $N_\phi$.  

If fragmentation occurs, then the number $N_\phi$ of detectable chameleons  is given in the production phase $-\tp < t \leq 0$ of the experiment by
\begin{eqnarray}
\frac{dN_\phi}{dt}
&=&
\fgam \Pgc - \Gdec N_\phi - \sfrag N_\phi^2 / V
\\
N_\phi(t)
&=&
\frac{(a^2-b^2) \sinh\left(\frac{a\sfrag(t+\tp)}{V}\right)}
{a\cosh\left(\frac{a\sfrag(t+\tp)}{V}\right) 
  + b \sinh\left(\frac{a\sfrag(t+\tp)}{V}\right) }
\qquad
\end{eqnarray}
where we have defined
\begin{eqnarray}
a
&=&
\sqrt{\left(\frac{\Gdec V}{2 \sfrag}\right)^2 
+ 
\frac{V \fgam \Pgc}{\sfrag}}\\
b
&=&
\Gdec V / (2 \sfrag).
\end{eqnarray}
In the afterglow phase $t>0$, the number $N_\phi(t)$ of chameleons and the time average of the afterglow signal $\faft(t) = \Gaft N_\phi(t)$ over the interval $t_0 < t < t_0 + \Delta t$  are given by
\begin{eqnarray}
N_\phi(t)
&=&
\frac{2 b N_\phi(0) e^{-\Gdec t}}
{2b + N_\phi(0)(1 - e^{-\Gdec t})}
\\
\left<\faft\right>
&=&
\Gaft N_\phi(0) / (\Gfm \Delta t) \nonumber \\
&& \times
\log\left[
1 
+ 
\frac{\Gfm (1-e^{-\Gdec \Delta t})}
{\Gdec + \Gfm(1 - e^{-\Gdec t_0})}
\right]\qquad
\label{e:faft_frag}
\end{eqnarray}
where the maximum fragmentation rate is $\Gfm = N_\phi(0) \sfrag / V$.

Given values of $g$, ${\mathcal N}$, $\bmat$, and $\bgam$, we can determine $\xiref$ from (\ref{e:xiref}), and $\meff$ as a function of the couplings from (\ref{e:meff_pwrlaw}).  We can compute the afterglow rate and the rate of decay to photons for these values of $\meff$, $\bgam$, and $\xiref$ using (\ref{e:Gaft}) and (\ref{e:Gdec}).  The resulting signal (\ref{e:faft_frag}) can be compared to our detection threshold, $3\sigma_\Gamma = 9.5$~Hz.  In this way, constraints on any power law chameleon may be obtained.

\begin{figure}[tb]
\includegraphics[angle=270,width=3.3in]{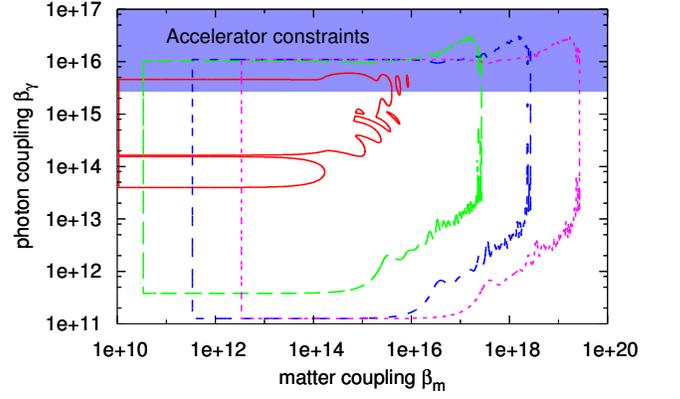}
\caption{Forecast constraints on the $\phi^4$ chameleon.  Solid (red), long dashed (green), medium dashed (blue), and short dashed (violet) correspond to $\lambda = 10^{-1}$, $10^{-2}$, $10^{-4}$, and $10^{-6}$, respectively; the shaded region shows the accelerator constraints of \cite{Brax_etal_2009}. \label{f:phi4}}
\end{figure}

Figure~\ref{f:phi4} shows the forecast GammeV--CHASE constraints on the $\phi^4$ chameleon, with $V(\phi) = \lambda \phi^4 /4!$.  The fragmentation interaction $\phi\phi \rightarrow \phi\phi\phi\phi$ has a cross section $\sfrag = \alpha_\mathrm{frag} \lambda^4 / \omega^2$, where $\alpha_\mathrm{frag} \sim 1$ is a numerical factor that cannot be written down in closed form for the four-body phase space of outgoing particles; we have approximated $\alpha_\mathrm{frag} = 1$.  We find that fragmentation prevents us from constraining the model with $\lambda = 1$ at any value of $\bgam$, and the models with $\lambda=0.1,0.01$ at low values of $\bgam$, while our constraints for smaller $\lambda$ are not limited by fragmentation.

The plots for $\lambda = 10^{-2}$, $10^{-4}$, and $10^{-6}$, shown as dashed lines in Fig.~\ref{f:phi4}, have a few distinctive features.  The upper and lower bounds in $\bgam$ are due to rapid chameleon decays and our detector sensitivity, respectively, as discussed earlier.  At low $\bmat$, we are limited by the requirement that chameleon particles be contained in our vacuum chamber, $\meff(\textrm{wall})>2.33$~eV.  Eq.~(\ref{e:meff_pwrlaw}) with $\rho_\mathrm{wall}=1$g/cm${}^3$ implies that $\meff = 1.6\times 10^{-3} \lambda^{1/6} \bmat^{1/3}$~eV.  For example, for $\lambda=10^{-2}$, containment requires that $\bmat > 3.4\times 10^{10}$.  Meanwhile, in the lower right corners of these three plots, corresponding to large $\bmat$ and small $\bgam$, the mass is essentially independent of $\bgam$ because $\bmat \rhom \gg \bgam \rhog$.  Thus the constraints in this region resemble the constraints in the lower right corner of Fig.~\ref{f:gammev-chase}, where at larger masses we need larger $\bgam$ in order to make up for the decline in the afterglow rate.  When both $\bmat$ and $\bgam$ are large, the chameleon mass also becomes large, pushing us into the regime where 3-point decay rates become important and our decay rate computation becomes unreliable.  Thus we have imposed a cutoff in the constraints at these masses, as discussed in Sec.~\ref{subsubsec:gammev-chase}.

The gaps and islands in the constraint plot for $\lambda=0.1$ demand further explanation.  We noted earlier that overlapping the constraints from multiple magnetic fields and multiple partitions in the magnetic field region helped to smooth out the sharp features seen in the afterglow and decay rates, Fig.~\ref{f:vary_mass}.  For example, a sudden dip in the afterglow due to one partition, if it occurred at a mass where the afterglow due to another partition was smooth and sufficiently large, would not show up as a feature in our constraint plot.  However, rapid fragmentation at $\lambda = 0.1$ reduces the afterglow signal from all three partitions, at all three $B$ values, so that they no longer overlap.  We obtain no constraints on the $\lambda=0.1$ models with the $B=5$~Tesla run, since that run probes low $\bgam$, where the afterglow rate is too small to produce an observable signal.  The gap in the constrained region at $\bgam = 1.6\times 10^{14}$ is due to poor overlap between the $B=1$~Tesla and $B=0.2$~Tesla runs.  The islands around $\bmat=10^{16}$ and $\bgam=10^{15}$ are due to bumps in $\Gaft$ as a function of $\meff$ in the $B=0.2$~Tesla run, in a region of $\bgam$ that is well beyond the reach of the $B=1$~Tesla run due to rapid fragmentation.

We note that GammeV--CHASE constraints on the $\phi^4$ chameleon will be at very different couplings from the constraints of laboratory fifth force searches \cite{Adelberger2007,Upadhye_Gubser_Khoury_2006}.  Furthermore, a $\phi^4$ chameleon massive enough to be contained in the vacuum chamber, $\meff(\rho_\mathrm{wall}) > \omega = 2.33$~eV, will have a mass in the Galaxy of $\meff(\textrm{Galaxy}) > 10^{-8}$~eV.  Since this is much larger than the plasma frequency $\omega_\mathrm{P} \approx 10^{-11}$~eV of gas in the galaxy, the chameleon-photon mixing angle will be strongly suppressed.  Thus, GammeV--CHASE will also probe $\phi^4$ chameleon models different from those ruled out by astrophysical dimming constraints.

\subsection{Chameleon dark energy}

\subsubsection{Inverse power law potentials}

First, we consider an inverse power law potential with a constant term, 
\begin{equation}
V(\phi)
=
\Mlam^4\left[ 1 + \kappa \left(\frac{\Mlam}{\phi}\right)^n\right],
\label{e:Vinv}
\end{equation}
where $n>0$, $\kappa>0$, and $\Mlam = \rho_\mathrm{de}^{1/4} = 2.4 \times 10^{-3}$~eV.  The model with $\kappa=1$ is the most economical, in the sense that the constant and power law terms both see the same mass scale, $\Mlam$; however, we allow for the possibility that $\kappa \neq 1$.  Eq. (\ref{e:xiref}) for ${\mathcal N}=-n$ implies that the phase shift at a wall reflection is
\begin{equation}
\xiref(n) = \frac{n\pi}{n+2}.
\end{equation}
The bulk field value $\phimin$, where $\partial \Veff(\phimin)/\partial \phi=0$, and the effective mass $\meff^2 = \partial^2\Veff(\phimin)/\partial \phi^2$ are given by
\begin{eqnarray}
\phimin
&=&
\Mlam \left(\frac{\kappa n \Mlam^3 \Mpl}{\bmat \rhom + \bgam \rhog}\right)\\
\meff^2
&=&
\kappa n (n+1) \Mlam^2
\left(\frac{\bmat \rhom + \bgam \rhog}
     {\kappa n \Mlam^3 \Mpl}\right)^\frac{n+2}{n+1}.
\end{eqnarray}
The mass scaling with density is given by $\eta = (n+2)/(2n+2)$, which is $3/4$ for $n=1$ and $2/3$ for $n=2$.  Only models with $n$ somewhat less than $1$ were accessible to GammeV, whereas all $n$ will be probed by GammeV--CHASE.

\begin{figure}[tb]
\begin{center}
\includegraphics[angle=270,width=3.3in]{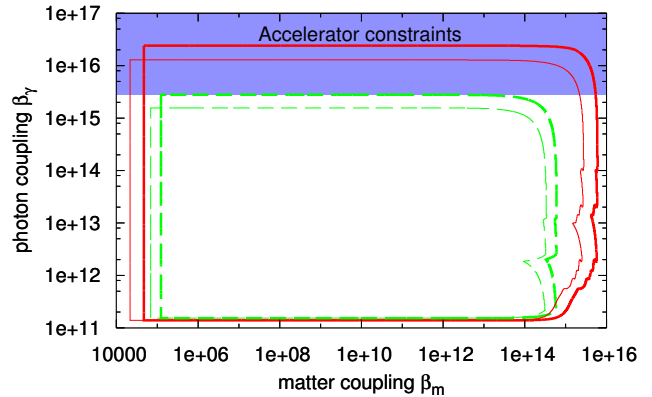}
\caption{Forecast GammeV--CHASE constraints on chameleon dark energy with an inverse power law potential (\ref{e:Vinv}).  Thick and thin lines refer to $\kappa = 1$ and $0.1$, respectively.  Solid (red) and dashed (green) lines refer to $n=1$ and $n=2$, respectively.  \label{f:de_pwr}}
\end{center}
\end{figure}
 
At the cosmological matter density $\rho_\mathrm{cos} \approx 2.5\times 10^{-30}$~g/cm${}^3$, the background field value for the $n=1$ model is $\phimin = 1.7\times 10^{-15} (\kappa/\bmat)^{1/2} \Mpl$, and the chameleon Compton wavelength is $\meff^{-1} \sim 100\textrm{ pc} \times (\kappa/\bmat)^{(2n+2)/(n+2)}$.  The Compton wavelength is tiny on cosmological scales, so the field remains close to the minimum of its potential.  Most of the chameleon energy density will come from the constant term $\Mlam^4$ in the potential.  For example, for $\kappa=1$, $\bmat = 10^{15}$, and $n=1$, the field-dependent term $\kappa(\Mlam/\phimin)^{n} \sim 10^{-8} \ll 1$, and this term is even smaller for larger $n$ and lower $\bmat$.  The field dependent term does not exceed the constant term until the matter density becomes of order $10^{-16}$~g/cm${}^3$, or about fourteen orders of magnitude greater than the current cosmological background density.  For all cosmological purposes, such a chameleon model behaves like a cosmological constant.  Evolution in its equation of state will be undetectable.

Since $\phi$ approaches zero at large densities, the total fractional change in the fine structure constant, between large densities and the cosmological background density $\rho_\mathrm{cos}$, is given by $\Delta \alpha_\mathrm{EM} / \alpha_\mathrm{EM} = \bgam \phimin(\rho_\mathrm{cos})/\Mpl$.  This is $\Delta \alpha_\mathrm{EM} / \alpha_\mathrm{EM} = 2.4\times 10^{-15} \bgam (\kappa/\bmat)^{1/2}$ for $n=1$, and even smaller for larger $n$.  Inverse power law models where $\bgam$ is sufficiently greater than $\bmat$ can be constrained by cosmological bounds on the variation in $\alpha_\mathrm{EM}$~\cite{Molaro_Reimers_Agafonova_Levshakov_2008,Murphy_Webb_Flambaum_2008}.  However, we note that chameleon models predict a density-dependence of $\alpha_\mathrm{EM}$, rather than a time-dependence.  Since astrophysical measurements of $\alpha_\mathrm{EM}$ are made in overdense regions of the universe, $\rhom > \rho_\mathrm{cos}$, the actual variation in $\alpha_\mathrm{EM}$ expected from chameleon theories will be smaller than the value given above.  Similarly, laboratory searches~\cite{Sortais_Bize_Abgrall_2001} for $\alpha_\mathrm{EM}$ variations carried out under conditions of constant density will not be sensitive to chameleons, though it may be possible to derive a constraint by comparing several $\alpha_\mathrm{EM}$ measurements such as \cite{Odom_Hanneke_DUrso_Gabrielse_2006} conducted at slightly different densities.  Laboratory and astrophysical constraints on chameleons from variations in $\alpha_\mathrm{EM}$ will require further analysis that is beyond the scope of this paper.

Photons from extragalactic sources, passing through the galactic magnetic field, could potentially oscillate into chameleon particles, causing the sources to appear dimmed, and possibly polarized~\cite{Carlson_Garretson_1994,Burrage_Davis_Shaw_2009}.  Constraints on astrophysical dimming have ruled out a wide range of photon couplings.  However, these constraints only apply to chameleons with masses much less than the Galactic plasma frequency $\omega_\mathrm{P} \approx 10^{-11}$~eV; larger masses strongly suppress chameleon-photon mixing in the low ($B \sim 10^{-10}$~Tesla) magnetic fields found in the Galaxy.  Assuming a Galactic matter density of $\sim 10^7 \rho_\mathrm{cos}$, the chameleon mass becomes greater than $\omega_\mathrm{P}$ for $\bmat > 10^{12}$ for $n=1$ and $\kappa=1$.  Larger values of $n$ and smaller values of $\kappa$ cause $\meff$ to exceed $\omega_\mathrm{P}$ for even lower values of $\bmat$; for $n=2$ and $\kappa = 0.1$, we find that $\bmat =10^{10}$ is sufficient.  Furthermore, astrophysical dimming constraints can be evaded altogether if the chameleon has a bare mass greater than $\omega_\mathrm{P}$, in addition to the mass it acquires from its matter coupling.

GammeV--CHASE will complement current constraints on inverse power law chameleons.  Figure~\ref{f:de_pwr} shows forecast GammeV--CHASE constraints for $n=1$, $2$, and $\kappa=1$, $0.1$.  The constraints are limited at low $\bgam$ by low afterglow rates, and at high $\bgam$ by rapid chameleon decays.  At low matter couplings, $\bmat \lesssim 10^4$, the chameleon mass in the walls of the vacuum chamber drops below their energy $\omega$, preventing the chamber from trapping the chameleon particles.  At large matter couplings, $\bmat > 10^{15}$, the constraints are limited by destructive interference due to large chameleon masses, and by the breakdown of our approximation $\Gdec(\textrm{2-point}) \approx \Gdec$.

\subsubsection{Exponential potentials}

\begin{figure}[tb]
\begin{center}
\includegraphics[angle=270,width=3.3in]{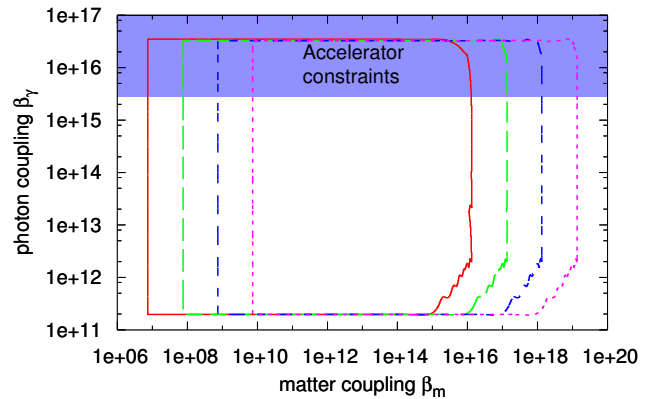}
\caption{Forecast GammeV--CHASE constraints on chameleon dark energy with an exponential potential.  Solid (red), long dashed (green), medium dashed (blue), and short dashed (violet) lines refer to $\kappa = 1$, $10^{-1}$, $10^{-2}$, and $10^{-3}$, respectively.  \label{f:de_exp}}
\end{center}
\end{figure}

Next, we study chameleon dark energy with an exponential potential,
\begin{equation}
V(\phi)
=
\Mlam^4 \left[1 + \exp\left(-\frac{\kappa\phi}{\Mlam}\right)\right],
\end{equation}
where, once again, $\Mlam = 2.4\times 10^{-3}$~eV, and $\kappa > 0$.  The constant term $\Mlam^4$ ensures that $\phi$ behaves as a dark energy; without it, the energy density of the field would scale with the background matter density, and accelerated expansion would not take place.  For this potential, we find, using the techniques of \cite{Brax_etal_2007},  that the phase shift at a wall reflection is $\xiref = \pi$.  

The bulk field value and the corresponding effective mass are given by
\begin{eqnarray}
\phimin
&=&
\frac{\Mlam}{\kappa} 
\log\left(\frac{\kappa\Mlam^3\Mpl}{\bmat\rhom+\bgam\rhog}\right)\\
\meff^2
&=&
\frac{\kappa\left(\bmat\rhom + \bgam\rhog\right)}{\Mlam\Mpl}.
\end{eqnarray}
The mass scales as the square root of the density, making exponential potentials inaccessible to GammeV, but well within the abilities of GammeV--CHASE.
 
The field value changes very little over a large range of densities.  The total variation in the $\log$ term in $\phimin$, over a range of densities from $\rho_\mathrm{cos}$ to laboratory densities $\sim 1$g/cm${}^3$, is less than $100$.  Thus the total change in $\phimin$ between these densities is of the order of $10^{-28}\kappa^{-1}\Mpl$.  Such a small change in $\phimin$ means that cosmological variations in $\alpha_\mathrm{EM}$ will not be observable for $\bgam \lesssim 10^{23}$, assuming that $\kappa$ is of order unity~\cite{Molaro_Reimers_Agafonova_Levshakov_2008,Murphy_Webb_Flambaum_2008}.  

At the cosmological density, the chameleon Compton wavelength is $\meff^{-1} = 4.7\times 10^{-6}\textrm{ pc} \times (\kappa\bmat)^{-1/2}$, so the field remains close to $\phimin(\rho_\mathrm{cos})$ on cosmological scales.  As with inverse power law chameleons in the previous section, the field-dependent term in the potential is small compared to the constant term, $\exp(-\kappa\phi/\Mlam) = 3.2\times 10^{-31} \bmat/\kappa \ll 1$, so the field behaves as a cosmological constant.  

For the exponential chameleon, constraints from astrophysical dimming do not overlap with constraints from GammeV--CHASE.  Any chameleon massive enough to be trapped in the GammeV chamber will have a mass greater than $10^{-11}$~eV at galactic densities, and dimming will be suppressed.  

Figure~\ref{f:de_exp} shows forecast GammeV--CHASE constraints on chameleon dark energy with an exponential potential.  As in the inverse power law case, our constraints are limited at low $\bgam$ by a low afterglow rate, at high $\bgam$ by rapid chameleon decays, at low $\bmat$ by chameleons too light to be trapped in our chamber, and at high $\bgam$ by destructive interference in chameleon-photon oscillation and the breakdown of our approximation that $\Gdec(\textrm{2-point}) \approx \Gdec$.  


\section{Conclusion}
\label{sec:conclusion}

GammeV and GammeV--CHASE are chameleon afterglow experiments which probe the coupling between photons and chameleon particles that are trapped inside a vacuum chamber.  Using the physics of chameleon-photon oscillation, we have computed the rate $\Gdec$ at which a chameleon converts to a photon inside the vacuum chamber of such an experiment, and the rate $\Gaft$ of production of photons which escape the chamber to reach an external detector.  We have computed these rates as functions of the chameleon-photon coupling $\bgam$, the effective chameleon mass $\meff(\textrm{chamber})$ inside the vacuum chamber, and the phase $\xiref$ that is introduced between chameleons and photons as they reflect from the walls of the vacuum chamber.  The decay and afterglow rates for the GammeV geometry are plotted as functions of $\meff$ in Fig.~\ref{f:vary_mass} and $\xiref$ in Fig.~\ref{f:vary_xiref}.  With these rates, we have computed the number of chameleons in the chamber and the expected afterglow flux in our detector, as a function of $\bgam$, $\meff(\textrm{chamber})$, and $\xiref$.

In a previous result, GammeV \cite{Chou_etal_2009} used the decay rate computed above, as well as the afterglow rate due to non-bouncing trajectories, to exclude a region of the chameleon parameter space.  Here, we have included bouncing trajectories in our afterglow calculation, resulting in constraints that are less conservative and that depend on the phase shift $\xiref$.  These model-independent constraints are shown in Fig.~\ref{f:gammev1}.  For $\xiref=0$ and small masses, the new constraints extend from half of the minimum photon coupling excluded by \cite{Chou_etal_2009} to twice the maximum coupling excluded by that reference.  For nonzero phases, additional suppression of the chameleon decay rate to photons allows us to push to even higher $\bgam$, excluding photon couplings ten times as high as those ruled out by \cite{Chou_etal_2009}.  However, since GammeV was only able to probe chameleon models whose masses scaled rapidly with density, even this new analysis does not allow us to constrain the most common types of chameleon potentials.

GammeV--CHASE is an improved version of GammeV that is expected to take data in the winter of 2009-2010.  In Fig.~\ref{f:gammev-chase} we forecast the model-independent chameleon constraints that will be achieved by GammeV--CHASE.  These constraints span many orders of magnitude in $\bgam$, bridging the gap between the constraints from GammeV and from particle accelerators.  We show that the improved control of PMT systematics will allow GammeV--CHASE to probe smaller couplings than GammeV.  Rapid PMT switching and multiple runs at lower magnetic fields will allow us to constrain $\bgam$ as high as $10^{16}$.  Our forecasts show that GammeV--CHASE will be able to probe a large range of couplings even at masses as high as the dark energy mass scale, $2.4\times 10^{-3}$~eV.  Furthermore, improvements to the pumping system allow GammeV--CHASE to probe many commonly used and well understood chameleon potentials, including quartic potentials, as well as inverse power law and exponential potentials that could explain the observed cosmic acceleration.

Quartic chameleons have been discussed extensively in the literature~\cite{Gubser_Khoury_2004,Upadhye_Gubser_Khoury_2006,Mota_Shaw_2006,Mota_Shaw_2007}. Torsion pendulum experiments such as \cite{Adelberger2007} have ruled out matter couplings up to $\bmat = 1$, but are insensitive to chameleons with stronger matter couplings.  It is precisely these chameleons which will be trapped in the GammeV--CHASE chamber.  We have shown that GammeV--CHASE will complement laboratory searches for quartic chameleons by probing strongly coupled chameleons, as shown in Fig.~\ref{f:phi4}.  For self-couplings a few orders of magnitude smaller than unity, constraints on quartic chameleons will span seven orders of magnitude in $\bmat$ and five orders of magnitude in $\bgam$, extending from the upper bound $\bgam \sim 10^{16}$ provided by particle accelerators down to $\bgam \sim 10^{11}$.  GammeV--CHASE is also complementary to searches for a varying fine structure constant.  If we extend a simple model of $\alpha_\mathrm{EM}$ variation such as \cite{Bekenstein1982} to a chameleon theory, by adding a $\phi^4$ potential and a Yukawa matter coupling with coupling constant proportional to mass, then constraints from GammeV--CHASE will be more powerful than those from laboratory or cosmological tests in a large portion of the parameter space.

Chameleon dark energy, with runaway potentials of the form $V(\phi) = M_\Lambda^4 f(\phi/M_\Lambda)$ and $M_\Lambda = 2.4\times 10^{-3}$~eV, have also been studied in the literature \cite{Brax_etal_2004}.  In such potentials, $f(\phi/M_\Lambda) \rightarrow 1$ as $\phi$ runs off to large values, corresponding to low matter densities such as the current cosmological background density.  We have applied our forecast GammeV--CHASE constraints to chameleon dark energy models with inverse power law potentials $V(\phi) = M_\Lambda^4 [ 1 + (M_\Lambda/\phi)^n]$ as well as exponential potentials $V(\phi) = M_\Lambda^4 [ 1 + \exp(-\phi/M_\Lambda)]$.  Figure~\ref{f:de_pwr} shows the constraints that GammeV--CHASE will be able to place on inverse power law chameleon dark energy.  For the simplest model, $n=1$, we will be able to probe matter couplings ranging from $\bmat \lesssim 10^{5}$ to $\bmat \lesssim 10^{16}$ and from $\bgam \gtrsim 10^{11}$ all the way up to the accelerator constraints, $\bgam \sim 10^{16}$.  At the largest matter ouplings, $\bmat > 10^{12}$, we will probe parameters inaccessible to astrophysical dimming constraints and bounds on variations in the fine structure constant.  GammeV--CHASE constraints on dark energy with an exponential potential, shown in Fig.~\ref{f:de_exp}, will be similarly powerful, covering matter couplings in the range $10^6 < \bmat < 10^{16}$.  Furthermore, this range of matter couplings is completely inaccessible to astrophysical dimming constraints, which apply to chameleons with low masses in the galaxy, and bounds from $\alpha_\mathrm{EM}$ variation, applicable to models with large variations in $\phi$.  Thus we have shown that GammeV--CHASE will probe large ranges of previously unexplored parameter space for the simplest models of chameleon dark energy.


\subsection*{Acknowledgments}
We are grateful to A. Baumbaugh, A. Chou, S. Gubser, C. Hogan, W. Hu, J. Khoury, A. Kusaka, P. O. Mazur, B. Odom, L. Reyes, A. Tolley, R. Tomlin, and W. Wester for many informative discussions.  This work was supported by the Kavli Institute for Cosmological Physics (KICP) at the University of Chicago through grants NSF PHY-0114422 and NSF PHY-0551142, as well as by the U.S. Department of Energy under contract No. DE-AC02-07CH11359.  JS thanks the Brinson Foundation for its generous support.  


\bibliography{gammevtheory}{}

\end{document}